\documentclass{ws-rv9x6}
\usepackage{subfigure}   
\usepackage{ws-rv-thm}   
\usepackage{bm}
\usepackage{ws-rv-van}   
\usepackage{caption}
\usepackage{subfigure}
\usepackage{commath}

\makeindex

\begin{document}

\chapter[Astrophysical constraints on strong modified gravity]{Astrophysical constraints on strong modified gravity}\label{ra_ch1}

\author[D. P\'erez and G. E. Romero]{Daniela P\'erez and  Gustavo E. Romero\footnote{Also at Facultad de Ciencias Astron\'omicas y Geof{\'\i}sicas, UNLP,\\ Paseo del Bosque s/n, 1900 La Plata, Buenos Aires, Argentina.}}
\address{Instituto Argentino de Radioastronom{\'\i}a, \\
(CCT - La Plata, CONICET - CICPBA),\\
C.C.5, 1894 Villa Elisa, \\
Buenos Aires, Argentina, \\
danielaperez@iar-conicet.gov.ar,\\
romero@iar-conicet.gov.ar}




\begin{abstract}
We offer a discussion on the strong field regime predictions of two families of theories that deviate from General Relativity in different aspects: $f(R)$-gravity and Scalar-Tensor-Vector Gravity (STVG). We discuss astrophysical effects in models based upon both matter and vacuum solutions of such theories. In particular, we analize neutron star structure and the constraints on the parameters of the theories introduced by the latest observations. We also review black hole solutions and several astrophysical consequences of them, including accretion disks and jets. Finally, we report on the implications of the detection of various gravitational wave events for these theories. 
\end{abstract}
\body

\section{General Relativity in the strong field domain: problems and challenges}

General Relativity (GR) is a theory of space, time, and gravitation formulated by Albert Einstein in 1915 \cite{Einstein1916}. In the theory spacetime is considered as an entity endowed with physical properties. Its physical geometry is represented by a continuous and differential 4-dimensional pseudo-Riemannian manifold $M$ with a metric field $g_{\mu\nu}$. The geometric properties of the manifold are related to the different matter fields existing in spacetime by Einstein's field equations:
\begin{eqnarray}
  R_{\mu\nu}-\frac{1}{2} R g_{\mu\nu}=\frac{8\pi G}{c^4} T_{\mu\nu},
  \label{EFE}
\end{eqnarray}

\noindent where $R_{\mu\nu}$ is the Ricci tensor obtained from the contraction of the Riemann curvature tensor, $R$ is the Ricci (or curvature) scalar, and $T_{\mu\nu}$ is the energy-momentum of all material fields. This is a set of ten nonlinear hyperbolic-elliptic partial differential equations in the coefficients of the metric field. Solving the equations for some distribution of energy and momentum, one can determine the free motion of test particles through the geodetic equation:
\begin{eqnarray}
  \frac{d^{2}x^{\lambda}}{ds^{2}}+\Gamma^{\lambda}_{\mu\nu}\frac{dx^{\mu}}{ds}\frac{dx^{\nu}}{ds}=0,
\label{motion}  
\end{eqnarray}

\noindent  where  $ds^2= g_{\mu\nu} dx^{\mu}dx^{\nu}$ is the spacetime interval and $\Gamma^{\lambda}_{\mu\nu}$ is the {\sl affine connection}  of the manifold:
 
\begin{equation}
\Gamma^{\lambda}_{\mu\nu}=\frac{1}{2}g^{\lambda\alpha}(\partial_{\mu}g_{\nu\alpha}+\partial_{\nu}g_{\mu\alpha}-\partial_{\alpha}g_{\mu\nu}).
\end{equation}

In GR, then, the effects of gravitation are a consequence of the curvature of spacetime. These effects exist in a non-local way, since curvature always vanishes on sufficiently small scales.

GR works wonderfully in the weak field regime. It passes all tests performed in the Solar system \cite{Will2014} and provides an adequate description of most astrophysical phenomena.  The theory correctly predicts the value for the perihelion advance of Mercury and the bending of light around the Sun. Gravitational Redshift, another classic prediction of the theory, has been successfully tested with different experiments. Shapiro delay was also confirmed with high confidence (a parameter $g = 1.000021 \pm 0.000023$, against a value of  $g = 1$ for GR).  Frame-dragging and the Geodetic effect have also been confirmed. The strong equivalence principle  has been tested to $h = 4.4 \times 10^{-4}$, with $h = 0$ in GR. Gravitational lensing has also confirmed GR to better than 1\%. More radically, the prediction of the existence of gravitational waves was spectacularly verified in 2015 with the detection of the first black hole binary merger event, dubbed  GW150914 \cite{abb+16a}.

Despite all these successes, GR is not free of problems. Singularities naturally appear in the theory for a number of spacetime models of both astrophysical \cite{Penrose1965} and cosmological importance \cite{Hawking-Penrose1970}.  In addition, black hole thermodynamics seems to suggest an incompatibility between GR and unitary evolution of quantum systems \cite{Giddings1995}. When applied to large scales, as those of galaxies and clusters of galaxies, GR requires the inclusion of mysterious dark matter to explain rotation curves and galaxy velocities. And when applied to the universe as a whole, a strange field of ``dark energy'' must be assumed to account for the observed accelerated expansion. Also, spacetime is expected to have quantum properties at the Planck scale, but GR is not renormalizable and therefore cannot be used to make meaningful physical predictions on such scales.

Not surprisingly, many attempts have been made at producing new spacetime theories that might overcome some of these problems. These theories are generally called ``modified gravity''. They should be almost identical to GR on the scales of the Solar System, where Einstein's theory is in accord with the experiments with exquisite accuracy. They might differ, however, in some not well-explored domains, such as large scales and in the strong gravity regime. In this article we shall explore some of the constraints imposed by current astrophysical observations upon some of these theories.

\section{Modified gravity: different approaches}

If a particle moves in spacetime departing from the expected geodetic trajectory, we face one out of three possibilities: 1) The trajectory actually is not a geodetic one and there are non-gravitational fields acting on the particle, 2) The trajectory is geodetic but there is matter not taken into account into our energy-momentum tensor, or 3) The trajectory is geodetic but the law that relates spacetime properties with energy-momentum is not correctly described by Eqs.~(\ref{EFE}). Situation 1) applies, for instance, to a charged particle affected by a magnetic field. Situation 2 requires a modification of the ontology accepted by our theory. This is the case when we keep Einstein's field equations untouched but  we introduce dark matter or dark fields to explain rotation galactic curves, gravitational lensing, or the accelerated expansion of the universe. Option 3 is one of ontological parsimony: in order to keep our ontology at a minimum we adopt a new prescription for the way spacetime interacts with other fields. This latter path demands modifications into the left side of Einstein's field equations, i.e. in the properties of spacetime. Such changes, of course, should be subtle enough as to yield the same predictions obtained from the original equations in those domains where the theory has passed stringent tests. The new solutions, however, might differ substantially from those of GR in the little known strong regime of high curvature or on cosmological scales. 

There are a number of ways to modify Eqs.~(\ref{EFE}) in order to achieve such effects. Einstein himself championed these attempts since 1917 till his death. His first modification consisted in introducing a cosmological term linear in the metric. This changes Eqs.~(\ref{EFE}) into:

\begin{eqnarray}
  R_{\mu\nu}-\frac{1}{2} R g_{\mu\nu} + \Lambda g_{\mu\nu}=\frac{8\pi G}{c^4} T_{\mu\nu}.
  \label{EFElambda}
\end{eqnarray}

\noindent Here, $\Lambda$ is the so-called cosmological constant. The effect of the new term is to allow for repulsive gravity over some scales. Einstein fine tuned $\Lambda$ to obtain a static (latter shown to be unstable) solution. Currently, such a term with a constant of value $\Lambda=1.11\times10^{-52}$ m$^{-2}$ is used in the standard cosmological model that includes cold dark matter (CDM), the $\Lambda$CDM model. In such a model the accelerated expansion is accounted for by the gravitational repulsion experienced by the cosmic fluid over some critical size. The same result can be obtained replacing the cosmological term on the left side of the equations by a positive vacuum energy density on the right side. Such energy density is called ``dark energy''. Notice that these are two different approaches that yield the same result: in the first case we modify the law of gravitation; in the second one, we add a dark field with a peculiar equation of state of the form $p=-\rho$.

A general way of introducing changes in the geometric sector of Eqs.~(\ref{EFE}) is to change the relativistic action. GR is obtained from the action:

\begin{equation}
	S[g]= \int {1 \over 2\kappa} R\, \sqrt{-g} \, \mathrm{d}^4x \label{Sg-fR}.
\end{equation}

\noindent This action can be generalized to

\begin{equation}
S[g]= \int {1 \over 2\kappa} f(R) \, \sqrt{-g} \, \mathrm{d}^4x,
\end{equation}

\noindent where $g$ is the determinant of the metric tensor and $f(R)$ is some function of the curvature (Ricci) scalar.

The generalized field equations are obtained by varying with respect to the metric. The variation of the determinant is

$$\delta\left(\sqrt{-g} \right)= -\frac{1}{2} \sqrt{-g}\, g_{\mu\nu} \delta g^{\mu\nu}.$$

\noindent The Ricci scalar is defined as

$$R = g^{\mu\nu} R_{\mu\nu}.$$

\noindent Therefore, its variation with respect to the inverse metric $g^{\mu\nu}$ is given by

\begin{eqnarray}
\delta R &=& R_{\mu\nu} \delta g^{\mu\nu} + g^{\mu\nu} \delta R_{\mu\nu}\nonumber\\
         &= & R_{\mu\nu} \delta g^{\mu\nu} + g^{\mu\nu}(\nabla_\rho \delta \Gamma^\rho_{\nu\mu} - \nabla_\nu \delta \Gamma^\rho_{\rho\mu}).
\end{eqnarray}

\noindent Since $\delta \Gamma^\lambda_{\mu\nu}$ is actually the difference of two connections, it should transform as a tensor. Therefore, it can be written as 

$$\delta \Gamma^\lambda_{\mu\nu}=\frac{1}{2}g^{\lambda a}\left(\nabla_\mu\delta g_{a\nu}+\nabla_\nu\delta g_{a\mu}-\nabla_a\delta g_{\mu\nu} \right),$$

\noindent and substituting in the equation above we get:

$$\delta R= R_{\mu\nu} \delta g^{\mu\nu}+g_{\mu\nu}\Box  \delta g^{\mu\nu}-\nabla_\mu \nabla_\nu \delta g^{\mu\nu}.$$

The variation in the action results in:

\begin{eqnarray}
\delta S[g]&=&\frac{1}{2\kappa}\int  \left(\delta f(R)\, \sqrt{-g}+f(R) \,\delta \sqrt{-g} \right)\, \mathrm{d}^4x \nonumber\\
           &= &\frac{1}{2\kappa}\int \left(F(R) \,\delta R \,\sqrt{-g}-\frac{1}{2} \sqrt{-g} \,g_{\mu\nu} \delta g^{\mu\nu} \,f(R)\right) \, \mathrm{d}^4x \nonumber \\
           &= &\frac{1}{2\kappa}\int \sqrt{-g}\,\left[F(R)(R_{\mu\nu} \delta g^{\mu\nu}+g_{\mu\nu}\Box  \delta g^{\mu\nu}-\nabla_\mu \nabla_\nu \delta g^{\mu\nu}) -\frac{1}{2} g_{\mu\nu} \,\delta g^{\mu\nu} f(R) \right]\, \mathrm{d}^4x, \nonumber
\end{eqnarray}

\noindent where $F(R)=\frac{\partial f(R)}{\partial R}$. Integrating by parts on the second and third terms we get

$$
\delta S[g]= \frac{1}{2\kappa}\int \sqrt{-g}\;\delta g^{\mu\nu} \left[F(R)R_{\mu\nu}-\frac{1}{2}g_{\mu\nu} f(R)+\left(g_{\mu\nu}\Box  -\nabla_\mu \nabla_\nu\right)F(R) \right]\, \mathrm{d}^4x. 
$$

By demanding that the action remains invariant under variations of the metric, i.e. $\delta S[g]=0$, we find the field equations in generic $f(R)$-gravity:

\begin{equation}
F(R)R_{\mu\nu}-\frac{1}{2}f(R)g_{\mu\nu}+\left[g_{\mu\nu} \Box -\nabla_\mu
\nabla_\nu \right]F(R) = \kappa T_{\mu\nu}, \label{eq-fR}
\end{equation}

\noindent where $T_{\mu\nu}$ is the energy-momentum tensor defined as

$$T_{\mu\nu}=-\frac{2}{\sqrt{-g}}\frac{\delta(\sqrt{-g}\; L_{\rm m})}{\delta g^{\mu\nu}},$$

\noindent and $L_{\rm m}$ is the matter Lagrangian. If $F(R)=1$, i.e. $f(R)=R$, we recover Einstein's theory.  

Equations (\ref{eq-fR}) are a system of non-linear partial differential equations of four order in the coefficients of the metric tensor field $g_{\mu\nu}$. A full description of $f(R)$-gravity can be found in the book by Capozziello and Faraoni \cite{cap+11}. An interesting feature is that the Ricci scalar $R$ and the trace of the energy-momentum tensor $T=T^{\mu}_{\nu}$ are related in a differential way. This implies that for some prescriptions of $f(R)$, the Ricci scalar can be different from zero even if $T=0$.

In the case of constant curvature $R=R_0$, Eqs.~(\ref{eq-fR}) in the absence of matter fields become:

\begin{eqnarray}
  R_{\mu\nu}=-\Lambda g_{\mu\nu},
\end{eqnarray}

\noindent where
\begin{eqnarray}
  \Lambda=\frac{f(R_0)}{f'(R_0)-1}.
\end{eqnarray}

\noindent We have, then, that cosmological solutions with accelerated expansion can be obtained in $f(R)$-gravity without adopting cosmological constant term or dark fields.

In GR, the effects of gravity are understood as the result of the curvature of spacetime. Such a curvature is described by the Riemann tensor, which consists of second order derivatives of the tensor metric field. A different approach to modified gravity consists in ascribing  to spacetime not only tensor fields, but also scalar and vector aspects. One of such attempts is known as Scalar-Tensor-Vector-Gravity (STVG) and has been presented by J. Moffat \cite{mof06}. 

In STVG theory, gravity is not only an interaction mediated by a tensor field, but by scalar and vector fields. The action of the full gravitational field is:
	\begin{equation}	
	S=S_{\mathrm{GR}}+S_{\phi}+S_{\mathrm{S}}+S_{\mathrm{M}},\label{Action}
	\end{equation}
\noindent where
\begin{equation}
S_{\mathrm{GR}}= \frac{1}{16 \pi G} \int d^4x \sqrt{-g}R,
\end{equation}
\begin{equation}	
S_{\phi}= - \omega \int d^4x \sqrt{-g} \left( \frac{1}{4} B^{\mu \nu} B_{\mu \nu} - \frac{1}{2} \mu^2 \phi^\mu \phi_\mu \right),
 \end{equation}
\begin{equation}
	\begin{aligned}
   S_{\mathrm{S}}&=  \int d^4x \sqrt{-g} \left[ \frac{1}{G^3} \left(\frac{1}{2} g^{\mu \nu} \nabla_\mu G \nabla_\nu G + V(G) \right) \right.+\\
      & \left. \frac{1}{G \mu^2} \left(\frac{1}{2} g^{\mu \nu} \nabla_\mu \mu \nabla_\nu \mu + V(\mu) \right)\right].
  \end{aligned}
\end{equation}
Here, $g_{\mu \nu}$ denotes the spacetime metric, $R$ is the Ricci scalar, $\nabla_{\mu}$ the covariant derivative, $\phi^{\mu}$ denotes a Proca-type massive vector field, $\mu$ is the mass and $B_{\mu \nu}=\partial_{\mu} \phi_{\nu} - \partial_{\nu} \phi_{\mu}$. $V(G)$ and $V(\mu)$ denote possible potentials for the scalar fields $G(x)$ and $\mu(x)$, respectively. We adopt units such that $c=1$. The term $S_{\mathrm{M}}$ refers to possible matter fields.\par

Varying the action with respect to $g^{\mu \nu}$ and doing some simplifications, the field equations result
$G_{\mu \nu} = 8\pi G \left( T^{\mathrm{M}}_{\mu \nu} + T^{\phi}_{\mu \nu} \right)$, where $G_{\mu \nu}$ denotes the Einstein tensor, and $T^{\mathrm{M}}_{\mu \nu}, T^{\phi}_{\mu \nu}$ are the matter and vector field energy-momentum tensors, respectively. The enhanced gravitational coupling is
$G=G_{\mathrm{N}}(1+\alpha)$,
where $G_{\mathrm{N}}$ denotes Newton's gravitational constant, and $\alpha$ a free parameter. Notice that STVG coincides with GR for $\alpha=0$.\par

Variation of the simplified action with respect to $\phi_{\mu}$ yields:
\begin{equation}
\label{vectoreq}
\nabla_{\nu} B^{\mu \nu} = -\frac{\sqrt{\alpha G_{\mathrm{N}}}}{\omega} J^{\mu},
\end{equation}
where $J^{\mu}$ denotes the four-current matter density, and the constant $\sqrt{\alpha G_{\mathrm{N}}}$ is determined to adjust the phenomenology. This vector field is completely absent in GR.\par

Certainly $f(R)$ and STVG are not the only families of modified gravity available. There are literally hundreds of alternative theories of gravitation, including multidimensional theories, torsion theories, bi-metric theories, theories of variable speed of light, and many, many more. An exhaustive discussion of the strong regime of all of them largely exceeds what can be contained in a single book, not to mention a single chapter. Hence, we opt to choose two of them, that we deem represent two different and rather natural generalizations of GR.  For more references on other approaches the reader can turn up to the mentioned book by Capozziello and Faraoni \cite{cap+11}.

How can we test the validity of theories such as STVG or $f(R)$-gravity? The answer is through their strong field effects. In this regime the fields behave differently from GR. Hence, studies of black holes, radiative effects in their surroundings, neutron stars, and other astrophysical objects where gravity is strong, are paramount to establish the validity of these theories well beyond the regime for which they were devised and originally applied. What remains of this chapter is devoted to discuss how the strong field regime astrophysics of compact objects changes if these theories were correct. Then, using the best available observational data we can impose some constraints of the validity domain of the theories.


\section{Neutron stars in modified gravity}

\subsection{Introduction}

Neutron stars are among the most compact astrophysical objects in the universe. They
are very dense stellar remnants where gravitational forces are balanced by
neutron degeneracy pressure. Their masses are in the range of $1.5$ $M_{\odot}$
to $2.2$ $M_{\odot}$ and radii typically of the order of 10
kilometers, reaching a density in the inner core of approximately $ 8
\times 10^{17}$ kg ${\rm m^{-3}}$, i.e., $10^{14}$ times the density of iron.
The intense gravitational field of neutron stars makes them ideal
objects to test current theories of gravitation in the strong field domain. 

The prediction of the existence of neutron stars was independent of astronomical observations. After the discovery of the neutron by Chadwick in 1932, Baade and Zwicky\cite{baa+34} were the first to suggest a new class of compact stars in which a core of degenerate neutrons could support the object against gravitational collapse\footnote{In the same work, they developed a theory for supernova explosions and proposed that these explosions could be the origin of cosmic rays.}. In 1939, Oppenheimer and Volkoff\cite{opp+39} and Tolman\cite{tol39} produced the first neutron star model, assuming the star as an ideal neutron gas. They showed the existence of an upper limit for the stellar mass of $0.75 M_{\odot}$ above which the star is not longer stable and collapses into a black hole.

After Second World War, the progress in the field of observational astronomy led to a series of discoveries that confirmed the theoretical predictions of previous decades. In 1967, a group of astronomers headed by Anthony Hewish detected astronomical objects that emitted regular radio pulses\cite{hew+67}. In the same year, Shklovsky\cite{shkl67} developed a detailed model to explain the radiation produced from Sco $X$-1, the first X-ray binary ever detected\cite{gia+62, san+66}. In his model, Shklovsky correctly established that the radiation was produced by the accretion of gas from a donor star onto a neutron star. The first binary pulsar was discovered by Hulse and Taylor in 1974\footnote{The pulsar mass was measured very precisely and it was found to be 1.44 $M_{\odot}$. The hypothesis of an ideal gas of neutrons for the interior of the star was ruled out, showing that the interactions between the nucleons needed to be considered\cite{cam07}.}. Pulsars are rapidly spinning neutron stars whose strong magnetic field produces conical beams of electromagnetic radiation. If the axis of rotation of the neutron star does not coincide with the magnetic axis, external observers see the beams whenever the magnetic axis points towards them as the star rotates. The pulses have the same period of the neutron star.

Given their clock-like regularity and their compact nature, pulsars offer a natural laboratory for studying the gravitational field. The analysis of the signals from pulsars in binaries provides information on the properties of these systems. In particular, pulsar timing techniques have proven to be extremely useful for estimating relativistic deviations from Keplerian motions in the case of binary pulsars with high velocities and strong gravitational fields. The measurement of the five post-Keplerian (PK) parameters\footnote{The five post-Keplerian (PK) parameters are: the rate of the periastron advance $\dot{\omega}$, the orbital period decay $\dot{P}_{b}$, the so-called relativistic $\gamma$ (the Einstein term corresponding to time dilatation and gravitational redshift), and the Shapiro delay term $r$ (range) and $s$ (shape).} made possible very precise test of GR and alternative theories of gravity.

The recent detection of gravitational wave (GW) signals from a stellar-mass binary system by the LIGO/VIRGO detectors\cite{abb+17b}, and the subsequent observation of a short $\gamma$-ray burst associated with the event made possible new approaches for studying the properties of neutron stars.  Using the data from GW170817 and general relativistic magnetohydrodynamics simulations (GRMHD), upper limits to the maximum gravitational mass, $M^{\rm sph}_{\rm max}$, of a non rotating, spherical neutron star were recently obtained\cite{rui+18}.

Neutron stars (NSs) have been investigated in the framework of some alternative theories of gravitation: for instance in Horndenski gravity\cite{mas+16}, Einstein-Gauss-Bonnet-Dilaton theory\cite{pan+11,kle+16}, and Chern-Simons Gravity\cite{yag+13}. In the following, we will focus on the main results for neutron star models in  $f(R)$-gravity and Scalar-Tensor-Vector Gravity, emphasizing  both the theoretical and observational predictions.

\subsection{Neutron star models in $f(R)$-gravity}

Compact objects have been largely studied in  $f(R)$-theories. The first works focused on the derivation and solution of the Tolman-Oppenheimer-Volkoff (TOV) equations that describe a spherically symmetric mass distribution in hydrostatic equilibrium. In order to solve the field equations, a perturbative approach was adopted, considering the $f(R)$ function as a perturbation of a GR background (see, for instance, Refs.\cite{coo+10,ara+11,del+12}).  Arapo\u{g}lu et al.\cite{ara+11} and Deliduman et al.\cite{del+12} used  the perturbative approach and adopted a  realistic Equation of State (EoS) for the matter distribution in R-squared and $R_{\mu \nu} R^{\mu \nu}$ gravities, respectively.

The mass-radius relation obtained by these authors for such $f(R)$ models allowed for larger masses of NSs than those currently estimated for the most massive known pulsars: $2.01 \pm 0.04  \ M_{\odot}$ for J0348+0432\cite{ant+13} and $1.928 \pm 0.017 \ M_{\odot}$ for J1614-2230\cite{dem+10}. Orellana and coworkers\cite{ore+13} have also studied the R-squared case using a polytropic approximation for the EoS, and also a more realistic one. The mass-radius relation obtained by these authors is consistent with the works previously mentioned: for the highest absolute values admitted for the free parameter of the theory, i.e., the $\alpha$ parameter, $f(R)$-gravity models predict higher masses of NSs than GR for every EoS.

An intriguing feature of the NSs models developed by Orellana et al. is a mass profile that in some regions decreases with the radius. In GR this effect could only be explained by means of a fluid of negative density. This is not the case in $f(R)$-gravity, where the coupling between the spacetime geometry and the matter could naturally give rise to such effects. This particular result, however, should be considered with caution as the authors clearly state, since it could be the consequence of the analytical representation of the EoS or the perturbative approach adopted.  

Later, it was pointed out by Yazadjiev and coworkers\cite{yaz+14} that the use of the perturbative method to investigate the strong field regime in $f(R)$-theories may lead to unphysical results\footnote{Recently, Bl\'azquez-Salcedo et al.\cite{bla+18} investigated the axial quasi-normal modes of the neutron star model in R-squared gravity developed by Yazadjiev and coworkers\cite{yaz+14}. }. In order to obtain self-consistent models of NSs, they suggested to solve the field equations simultaneously, assuming appropriate boundary conditions.  
This approach was then applied by some authors\cite{ast+15,cap+16,apa+16}. 

The internal structure of NSs was also explored using the Palatini formalism\cite{kai+07}. In the Palatini formalism the metric and the connection are a priori considered independent geometrical objects\footnote{In $f(R)$-gravity with a chameleon mechanism, Liu et al.\cite{liu+18} found constraints for a general $f(R)$ functions using observations of the pulsars PSR J0348+0432 and PSR J1738+0333. This restriction, however, is weaker than the one derived from the solar systems observations.}. The advantage of this formalism is that the field equations have derivatives of the metric up to second order, as in GR. Though the apparent mathematical simplification of the field equations, NSs models in this formalism present serious shortcomings as shown by Barause et al. in a series of two works\cite{bar+08,barf+08}. For a polytropic equation of state, the authors demonstrated the appearance of divergences in the curvature invariants near the surface of the star, indicating that the origin of the singularity is related to the intrinsic features of Palatini $f(R)$-gravity\cite{barf+08}. For a realistic EoS, and choosing the $f(R)$ function as $f(R) = R + \alpha R^{2}$, Barause and coworkers found that the radial profile of the mass parameter develops bumps when rapid changes in the derivatives of the EoS occur. 

Motivated by the later result, Teppa Pannia et al.\cite{tep+17} aimed to investigate whether the non-smoothness of the mass parameter was rooted to the nature of f(R)-gravity in the Palatini formalism, or was an effect of the particular EoS chosen. 

The method used was to calculate the structure of a star in the Palatini formalism in $R$-squared gravity for two EoS: first an EoS similar to the one employed by Barause et al.\cite{bar+08}; second, an EoS based on the connection of multiple polytropes that allows to control the derivatives of the EoS using a set of parameters. The results found were in accordance with those of Barause et al.\cite{bar+08,barf+08}:  for both EoS a) the maximum masses were lower than in GR , b) the mass profile displays regions where $dm/d\rho < 0 $.

In conclusion, these investigations strongly suggest that the odd features in NSs models in Palatini $f(R)$-gravity do not lay in the characteristics of the EoS nor in the particular mathematical method employ to solve the field equations, but are inherent to the nature of the theory\footnote{More complex models of NSs in $f(R)$-gravity have been developed which include the fluid anisotropy\cite{fol18}, rotation\cite{sta+14,yaz+15}, magnetic fields\cite{che+13,ast+15b,bak+16}, and different $f(R)$-functions and formalisms\cite{ala+13,liu+18}.}.

 \subsection{Neutron stars in Scalar-Tensor-Vector Gravity}
 
 Contrary to $f(R)$-gravity, neutron star models in STVG have just started to be explored. Currently, there is only one work in the scientific literature on this issue by Lopez Armengol and Romero\cite{lop+17}. These authors derived the modified TOV equation assuming a static, spherically symmetric geometry for the spacetime; the stellar matter content was modeled with a static, spherically symmetric, perfect fluid, energy momentum tensor.
 
 Four distinct neutron EoS were considered: POLY\cite{sil+04}, SLy\cite{dou+01}, FPS\cite{pan+89} and BSK21\cite{gor+10,pea+11,pea+12}. The first EoS is mathematically simple and well-behaved. The purpose of employing POLY to construct neutron star models was that if any particular feature arises in the model, it would probably be an effect of STVG. Conversely, SLy, FPS  and BSK21 are realistic EoS.
 
 When integrating numerically the equations, particular attention is needed for the free parameter $\alpha$ of the theory. The parameter plays a fundamental role since it mediates both the gravitational repulsion and enhanced gravitational attraction, being its value dependent on the mass source of the gravitational field. Moffat\footnote{The restriction on $\alpha$ was imposed in order to find agreement of STVG predictions with the perihelion advance of Mercury.} determined for solar mass sources an upper limit given by\cite{mof06}:
 \begin{equation}\label{moffat-rest}
 \alpha_{\odot} << \frac{1.5 \times 10^{5} \ c^{2}}{G_{\rm N}}\frac{1}{M_{\odot}} \ \ {\rm cm}. 
 \end{equation}
Here, $G_{\rm N}$ stands for Newton's gravitational constant, and $c$ denotes the speed of light. Since neutron stars have few solar masses, the values of $\alpha$ were chosen to satisfy inequality (\ref{moffat-rest}). It was also taken into account  that inside the star, $\alpha_{\rm NS}$ would depend on the mass of each $r$-shell. A linear ad hoc prescription was defined to sample different values of $\alpha$:
\begin{equation}
\alpha_{\rm NS} = \gamma  \frac{1.5 \times 10^{5} \ c^{2}}{G_{\rm N}}\frac{1}{M_{\odot}} \left(\frac{M(r)}{M_{\odot}}\right) \ \ {\rm cm},
\end{equation}
where $M(r)$ is the mass of the neutron star up to the $r$-shell, and $\gamma$ is a normalized factor $\gamma \in [0;1)$. 

One of the main findings of Lopez Armengol and Romero\cite{lop+17} is that neutron star models in STVG admit higher total masses than in GR. This result deserves particular attention since recent determinations of neutron star masses defy GR limits\cite{ant+13,kiz+13,oze+12,dem+10}. The authors could also set a more restrictive upper limit for the parameter $\alpha$ by imposing within their model, neutron stars with realistic masses (in accordance with astronomical observations), and monotonically decreasing density profiles. The new restriction for the $\alpha$ parameter for stellar mass sources is:
\begin{equation}
 \alpha < 10^{-2} \frac{1.5 \times 10^{5} \ c^{2}}{G_{\rm N}}\frac{1}{M_{\odot}} \\ {\rm cm}. 
\end{equation}

There are many issues that remain to be explored in relation to neutron stars in STVG; for instance, solutions that take into account the rotation of the star, the scalar field contributions, stability analysis, and quasinormal modes of spherically symmetric solutions. STVG, thus, seems to be a rich field of research for us in the future.


\section{Black holes in modified gravity}

Black holes constitute the most extreme manifestation of gravity. These objects are spacetime regions causally disconnected from the rest of the universe by an event horizon. Black holes were first postulated theoretically\cite{sch16}, and half a century later their astrophysical manifestation started to be detected\cite{haz+63,web+72,bol72}. There is overwhelming astronomical evidence that supports their existence. The latest and most direct proof is the detection of gravitational waves produced by the merger of binary systems of black holes\cite{abb+16a,abb+16c,abb+17a,abb+17e,abb+17a1}.

In the light of the recent discoveries, any viable alternative theory of gravitation should admit black hole solutions. This is the case of $f(R)$-gravity and STVG that now we will proceed to analyze. 

\subsection{Black holes in $f(R)$-gravity}\label{bh}

There is an extensive literature on black hole solutions in $f(R)$-gravity; during the course of a decade, almost every year new solutions have been reported. This situation is in stark contrast with GR. Because of the uniqueness theorem, we know that the only stable stationary asymptotically black hole solutions  in GR are within the Kerr-Newman family\cite{isr67,isr68,car71,hawk72,rob75}. In $f(R)$-gravity, however, the Birkoff theorem does not hold\cite{sot+10,far10}, and is still an open question whether ``hairy black'' hole solutions for a non constant Ricci scalar exist in $f(R)$-gravity\cite{can+16,can18}.

Black holes geometries have been found in $f(R)$-gravity both in the metric and Palatini formalisms\footnote{For black hole solutions in Palatini formalism see, for instance, Refs.\cite{olm+11,olm+12a,olm+12b,baz+14,Olm+15a}.}. As clearly stated by Sotiriou and Faraoni\cite{sot+10}, all black holes solutions of GR (with a cosmological constant) will also be solutions of $f(R)$ in both formalisms (see also Refs.\cite{psa+08,bar+08a}). In the Palatini formalism, they will comprise the complete set of black hole solutions of the theory.  In the metric formalism, the Birkoff theorem does not hold, thus, other black hole solutions can in principle exist.

In $f(R)$-gravity, the Ricci scalar can depend both on space and time. Given the extreme complexity of the field equations, a simplifying assumption is to consider the Ricci scalar constant $R = R_{0}$. Several black hole solutions were determined under this hypothesis: static spherically symmetric solutions\cite{del+09,del+11}, charged solutions\cite{moo+11} (see also Ref.\cite{hen+12}) and  the corresponding generalization in Kerr-Newman spacetimes\cite{cem+11}; static spherically symmetric solutions coupled to linear and non-linear electromagnetic fields\cite{maz+12}, and also minimally coupled to a non-linear Yang-Mills field\cite{maz+12b}; higher dimensional ($d \ge 4$) charged black holes have also been explored\cite{she12}.

Without assuming $R =R_{0}$, Perez Bergliaffa and Chifarelli de Oliveira Nunes studied the necessary conditions for an $f(R)$-theory to have static spherically symmetric black hole solutions\cite{ber+11}. Following the ``near horizon test'' introduced by the previous authors, Mazharimousavi and coworkers\cite{maz+13} derived the necessary conditions for the existence of Reisnner-Nordstr\"{o}m  black holes in $f(R)$-gravity.  

A formalism for the generation of spherically symmetric metrics in d-dimensions both in vacuum and in the presence of matter sources was given by Amaribi et al.\cite{ami+16}.  Gao and Shen\cite{gao+16} also proposed a new method to find exact solutions for static, spherically symmetric spacetimes in this theory\footnote{Further black hole solutions were studied in $f(R)$ with conformal anomaly\cite{hen+11}; regular $f(R)$-black holes solutions have also been explored (see for instance Ref. \cite{rod+16}). Stability analysis for different class of $f(R)$-black hole solutions can be found in Ref. \cite{myu+11,moo+11b,myu11}. }.

In what follows, we will describe the main properties of the $f(R)$-Kerr black hole with constant Ricci scalar since such spacetime has been employed to constrain the parameters of some class of $f(R)$-theories (see Section \ref{accretion-disks}).

The axisymmetric, stationary, and constant Ricci scalar geometry that describes a black hole with mass, electric charge, and angular momentum was found by Carter\cite{car73}. This geometry was used to study $f(R)$ black holes by Cembranos and coworkers\cite{cem+11}. The line element takes the form\footnote{The line element for a $f(R)$-Schwarzschild black hole can be obtained by setting $a = 0$.} (we have set $Q = 0$):
\begin{eqnarray}\label{ds}
ds^{2} & = & \frac{\rho^{2}}{\Delta_{\rm r}} dr^{2} +   \frac{\rho^{2}}{\Delta_{\rm \theta}} d\theta^{2}
+  \frac{\Delta_{\rm \theta} {\sin{\theta}}^{2}}{\rho^{2}}\left[a \frac{c \ dt}{\Xi}-\left(r^{2} + a^{2}\right) \frac{d\phi}{\Xi}\right]^{2}\nonumber \\
& - & \frac{\Delta_{\rm r}}{\rho^{2}}\left(\frac{c \ dt}{\Xi} - a \  {\sin{\theta}}^{2} \frac{d\phi}{\Xi} \right),
\end{eqnarray}
where,
\begin{eqnarray}
\Delta_{\rm r} & = & \left(r^{2}+a^{2}\right)\left(1 - \frac{R_{0}}{12} r^{2}\right) - \frac{2 G M r}{c^{2}},\\
\rho^{2} & = & r^{2} + a^{2} {\cos{\theta}}^{2},\\
\Xi & = & 1 + \frac{R_{0}}{12} a^{2}.  
\end{eqnarray}
Here $M$ and $a$ denote the mass and angular momentum of the black hole, respectively. If $R_{0} \rightarrow 0$, Eq.~(\ref{ds}) represents the spacetime metric in GR as expected\footnote{Calz\`{a}, Rinaldi and Sebastiani\cite{cal+18} studied spherically symmetric solutions in vacuum for a special class of $f(R)$ functions that satisfy: $f(R_{0}) = 0$, and $df(R_{0})/dR = 0$. Some of the metrics obtained represent spherically symmetric black holes. Under a specific choice of the values of some parameters, the $f(R)$-Schwarzschild black hole solution here presented  is recovered.}.

The relation between the Ricci scalar $R_{0}$ and the $f(R)$ function can be derived from the field equations of the theory in the metric formalism:
\begin{equation}\label{eq3}
R_{\mu \nu}\left(1 + f'(R)\right) - \frac{1}{2} g_{\mu \nu} \left(R + f(R)\right) + \left(\nabla_{\mu} \nabla_{\nu} - g_{\mu \nu} \Box\right) f'(R) + \frac{16 \pi G}{c^{4}} T_{\mu \nu} = 0,  
\end{equation}
where $R_{\mu \nu}$ is the Ricci Tensor, $\Box \equiv \nabla_{\beta} \nabla^{\beta}$, $f'(R) = df(R)/dR$, and $T_{\mu \nu}$ is the energy-momentum tensor. If we take the trace of the latter equation, we obtain:
\begin{equation}\label{eq5}
R \left(1+f'(R)\right) - 2 \left(R+f(R)\right) -3 \Box f'(R) + \frac{16 \pi G}{c^{4}} T = 0.  
\end{equation}
In the case of constant Ricci scalar $R_{0}$ without matter sources, from Eq.~(\ref{eq5}) we get the relation we were looking for:
\begin{equation}\label{eq7}
R_{0} = \frac{2 \ f(R_{0})}{f'(R_{0}) -1}.  
\end{equation} 

It also should be noticed that if $R = R_{0}$ and $T_{\mu \nu} = 0$, Eqs.~(\ref{eq3}) can be re-cast as:
\begin{equation}
R_{\mu \nu} = \Lambda g_{\mu \nu}, 
\end{equation}
and,
\begin{equation}\label{eq6}
\Lambda =  \frac{ f(R_{0})}{f'(R_{0}) -1}.
\end{equation}
From relations (\ref{eq7}) and (\ref{eq6}) we see that $f(R)$ with constant Ricci scalar and no matter sources is formally equivalent to GR with a cosmological constant. This equivalence, however, is not physical as some authors have recently confused\cite{can+16}.

The location of the event horizon as a function of the radial coordinate is obtained by setting $1/g_{\rm rr} = 0$:
\begin{equation}
 \Delta_{\rm r}  =  \left(r^{2}+a^{2}\right)\left(1 - \frac{R_{0}}{12} r^{2}\right) - \frac{2 G M r}{c^{2}}.
 \end{equation}

For a nearly maximally rotating black hole $a = 0.99$, such as Cygnus X1\cite{gou+11}, the existence (or absence) of horizons depends on the value of the Ricci Scalar\footnote{In what follows, we express the values of  Ricci scalar as a dimensionless quantity: $\mathsf{R_0} \equiv R_{0} r^{2}_{\rm g}$, where $r_{\rm g} = G M/c^{2}$.} $R_{0}$. If $\mathsf{R_0} \in (0,0.6]$, there are 3 event horizons: the inner and outer horizons of the black hole and a cosmological horizon;  for $\mathsf{R_0}>0.6$ there is  a cosmological horizon that becomes smaller for larger values of $\mathsf{R_0}$. If $\mathsf{R_0} \in (-0.13,0)$ there are 2 event horizons. For $\mathsf{R_0} \leq -0.13$ naked singularities occur\cite{per+13}. Black hole solutions, thus, occur for $\mathsf{R_ 0} \in (-0.13,0.6]$. 

\subsection{Black holes in Scalar-Tensor-Vector Gravity}\label{bh-stvg}

The only known black hole solutions in STVG were found by Moffat\cite{mof+15a}. These represent static, spherically symmetric and also stationary, axially symmetric black holes, and were derived under the following assumptions:
\begin{itemize}
\item The mass $m_{\phi}$ of the vector field $\phi^{\mu}$ was neglected because its effects manifest at kiloparsec scales from the source\footnote{The mass of the field $\phi^{\mu}$ determined from galaxy rotation curves and galactic cluster dynamics\cite{mof+13,mof+14,mof+15} is $m_{\phi} = 0.042$ kpc$^{-1}$.}.
\item $G$ is a constant that depends on the parameter $\alpha$ \cite{mof06}:
\begin{equation}
G = G_{\rm N} \left(1 + \alpha \right),
\end{equation} 
where $G_{\rm N}$ denotes Newton's gravitational constant, and $\alpha$ is a free dimensionless parameter.
\end{itemize}

Given these hypothesis, and after solving the STVG field equations, the line element of the spacetime metric of a black hole of mass $M$ and angular momentum $J = a M$ in STVG theory is \cite{mof+15a}:
\begin{eqnarray}\label{BH-metric}
ds^{2} & = &  -c^{2} \left(\Delta - a^{2} \sin^{2}\theta\right) \frac{dt^{2}}{\rho^{2}} + \frac{\rho^{2}}{\Delta} dr^{2} + \rho^{2} d\theta^{2}\nonumber \\
 &+&  \frac{2 a c \sin^{2}{\theta}}{\rho^{2}}\left[(r^{2}+a^{2})-\Delta\right] dt d\phi \nonumber \\
&+ & \left[(r^{2}+a^{2})^{2}-\Delta a^{2} \sin^{2}\theta\right] \frac{\sin^{2}\theta}{\rho^{2}} d\phi^{2},\\
\Delta & = & r^{2} - \frac{2 G M r}{c^{2}}+ a^{2} + \frac{\alpha G_{N} G M^{2}}{c^{4}}\label{Delta} \\
& = & r^{2} - \frac{2 G_{N} (1+\alpha) M r}{c^{2}}+ a^{2} + \frac{\alpha (1+\alpha) G_{N}^{2} M^{2}}{c^{4}},\nonumber\\
\rho^{2} & = & r^{2} + a^{2} \cos^{2}\theta.
\end{eqnarray}
The metric above reduces to the Kerr metric in GR when $\alpha = 0$. By setting $a = 0$ in Eq. \eqref{BH-metric}, we recover the metric of a Schwarzschild STVG black hole. 

The radius of the inner $r_{-}$ and outer $r_{+}$ event horizons are determined by the roots of $\Delta = 0$:
\begin{equation}\label{hori-Kerr}
r_{\pm} = \frac{G_{N} (1+\alpha)M}{c^{2}}\left\{1 \pm \sqrt{1- \frac{a^{2} c^{4}}{G_{N}^{2} (1+\alpha) M^{2}} - \frac{\alpha}{(1+\alpha)}}\right\}. 
\end{equation}
If we set $\alpha =  0$ in the latter expression, we obtain the formula for the inner and outer horizons of a Kerr black hole in GR. Inspection of Eq. \eqref{hori-Kerr} also reveals that for $\alpha > 0$ the outer horizon of a Kerr black hole in STVG is larger that the corresponding one in GR.

The radial coordinate of the ergosphere is determined by the roots of $g_{tt} =0$:
\begin{equation}
r_{E} = \frac{G_{\rm N} M \left(1+\alpha\right)}{c^{2}}\left[1 \pm \sqrt{1 - \frac{a^{2} \cos^{2}{\theta} \ c^{4}}{{G_{N}}^{2} \left(1+\alpha\right)^{2} M^{2} }-\frac{\alpha}{1+\alpha}}\right].  
\end{equation}
We see that the ergosphere grows in size as the parameter $\alpha$ increases. 

Astrophysical black holes strongly interact with the surrounding media giving rise to a wide variety of high energy phenomena; by studying the particles and radiation that is produced in these sources, the properties of the black hole spacetime can be inferred, and thus the underlying theory of gravitation can be put to the test.

In the coming two sections we provide a brief account of the salient features of accretion disks and relativistic jets around black holes in $f(R)$-gravity and STVG.



\section{Accretion disks}\label{accretion-disks}

The process of matter with angular momentum falling into a black hole may lead to the formation of an accretion disk.  The matter rotating around the black hole loses angular momentum because of the friction between adjacent layers and spirals inwards; in the process the kinetic energy of the plasma increases and the disk heats up emitting thermal energy.

The characterization of the nature of the turbulence in the disk, and hence of the dissipation mechanisms constitutes the main problem for the formulation of a consistent theory of accretion disks. Some simplifying assumptions, however, can be made in order to construct realistic models of disks. Next, we shall consider accretion disks in steady state where the accretion rate is an external parameter, and the turbulence is characterized by a single parameter ``$\alpha$'' which was first introduced by Shakura\cite{sha72} and Shakura and Sunyaev\cite{sha+73}.

Novikov, Page, and Thorne\cite{nov+73,pag+74}  generalized the latter model to include the strong gravitational fields effects. They assumed the background spacetime geometry to be stationary, axially symmetric, asymptotically flat,  reflection-symmetric with respect to the equatorial plane, and the self-gravity of the disk was considered negligible. The central plane of the disk is located at the equatorial plane of the spacetime geometry, and since the disk is assumed to be thin\footnote{$\Delta z = 2 h << r$, being $z$ the height above the equatorial plane, and $h$ is the thickness of the disk at radius $r$.} the metric coefficients  only depend on the radial coordinate $r$.

From the relativistic equations for the conservation of mass, energy, and angular momentum,  Page and Thorne\cite{pag+74} obtained three fundamental equations for the time-averaged radial structure of the disk\footnote{The average is taken over a time interval $\Delta t$ during which it is assumed that the external geometry of the hole is modified negligibly, but the accretion of matter for any radius $r$ is large compared with the typical mass enclosed in a ring of thickness $r$.}. In particular, they provided an expression for the heat emitted by the accretion disk that reads:
\begin{equation}\label{heat}
Q(r) = \frac{\dot{M}}{4 \ \pi \sqrt{-g}}\frac{\Omega_{,r}}{\left(E^{\dagger} - \Omega L^{\dagger}\right)^{2}} \int^{r}_{r_{\rm isco}} \left(E^{\dagger} - \Omega L^{\dagger}\right) \ L^{\dagger}_{,r} \ dr.
\end{equation}
Here, $\dot{M}$ stands for the mass accretion rate,  $g$ is the metric determinant, and $r_{\rm isco}$ denotes the radius of the innermost stable circular orbit. The expressions for the specific energy $E^{\dagger}$, specific angular momentum $L^{\dagger}$, and angular velocity $\Omega$ of the particles that move on equatorial geodesic orbits around the black hole are:
\begin{eqnarray}
E^{\dagger}(r) & \equiv  &  - u_{t}(r), \label{ene}\\
L^{\dagger}(r) & \equiv & u_{\phi}(r), \label{mom}\\
\Omega(r) & \equiv & \frac{u^{\phi}}{u^{t}} \label{ang},
\end{eqnarray}
where $u_{t}(r)$ and $u_{\phi}(r)$ are the $t$ and $\phi$ components of the four-velocity of the particle, respectively. Formulas (\ref{ene}), (\ref{mom}), and (\ref{ang}) can also be written in terms of the metric coefficients (see Harko et al.\cite{har+09} or P\'erez et al.\cite{per+13} for the corresponding expressions). 

 In order to compute the heat emitted by a thin accretion disk around a black hole, we first have to study the circular orbits of massive particles in such spacetime geometry. There are several papers devoted to the investigation of geodesics in both $f(R)$-gravity and STVG. In particular, the analysis of stable circular orbits of massive particles in $f(R)$-Schwarzschild and $f(R)$-Kerr black holes\footnote{Since $f(R)$ with constant Ricci scalar and no matter sources is formally equivalent to GR with a cosmological constant (see Section \ref{bh}), the results of P\'erez et al. are in accordance with the  the analysis of circular orbits in Schwarzschild-de Sitter, Schwarzschild-anti- de Sitter done by Stuchl{\'i}k and Slan\'y\cite{stu+99} and Rezzolla et al.\cite{rez+03}, and also in Kerr-de Sitter and Kerr-anti-de Sitter performed by  Stuchl{\'i}k and Slan\'y\cite{stu+04}, and Slan\'y and coworkers\cite{sla+13}.} was performed by P\'erez and coworkers\cite{per+13}. The existence and location of stable circular orbits depends on the value of the Ricci scalar $\mathsf{R_0}$. For both Schwarzschild and Kerr black holes in $f(R)$-gravity, if $\mathsf{R_0} > 0$ there is a minimum ($r_{\rm isco}$) and maximum ($r_{\rm osco}$) radius for stable circular orbits; for $r > r_{\rm osco}$ there are no stable circular orbits in such spacetime.  On the contrary, for $\mathsf{R_0} < 0$ stable circular orbits begin from a minimum radius and extend to the rest of the spacetime. In Table \ref{table1}, we show the values of the Ricci scalar for which stable circular orbits are possible (second column), and the corresponding $r_{\rm isco}$ and $r_{\rm osco}$ (columns 3 and 4, respectively) for $f(R)$-Schwarzschild and $f(R)$-Kerr spacetimes ($ a = 0.99$). The formulas from which these results were obtained can be found in\cite{per+13} and references therein.
 
 \begin{table}[ht]
\tbl{Values of the innermost and outermost stable circular orbits for $f(R)$-Schwarzschild and $f(R)$-Kerr spacetimes ($ a = 0.99$). The second column indicates the values of Ricci scalar $\mathsf{R_0}$ for which stable circular orbits are possible.}
{\begin{tabular}{@{}cccc@{}} \toprule 
Spacetime &   $\mathsf{R_0}$ & $r_{\rm isco}$ & $r_{\rm osco}$ \\  \colrule
$f(R)$ & $\mathsf{R_0} \in (0, 2.85 \times 10^{-3})$ &  $r_{\rm isco} \in (6, 7.5)$ &  $r_{\rm osco} \in (7.5, +\infty)$\\
Schwarzschild &  $\mathsf{R_0} < 0$  & $r_{\rm isco} \in (3.75, +\infty)$ & $-$   \\ \botrule
$f(R)$ & $\mathsf{R_0} \in (0, 1.45 \times 10^{-1})$ &  $r_{\rm isco} \in (1.4545, 2)$ &  $r_{\rm osco} \in (2, +\infty)$\\
Kerr &  $\mathsf{R_0} \in (-0.13, 0)$ & $r_{\rm isco} \in (1, 1.4545)$ & $-$\\ \botrule
\end{tabular}
}
\label{table1}
\end{table}
 
The existence and location of stable circular orbits in Schwarzschild and Kerr black holes in STVG were studied in Ref.\cite{per+17}. The range of values adopted for the parameter $\alpha$ were chosen in order to model accretion disks around stellar and supermassive black holes. For stellar mass sources, $\alpha < 0.1$ was taken in accordance to the work of Lopez Armengol and Romero\cite{lop+17}. In the case of supermassive black holes  ($10^{7} M_{\odot} \le M \le 10^{9} M_{\odot} $), the values of $\alpha$ were in the range $\alpha \in (0.03, 2.47)$. The lower limit for $\alpha$ was set in agreement with Moffat et al.\cite{mof+08} that predicted such a value for globular clusters, while the upper limit for the parameter was calculated by fitting rotation curves of dwarf galaxies ($1.9 \times 10^{9} M_{\odot} \le 3.4 \times 10^{10} M_{\odot}$) by Brownstein and coworkers\cite{bro+06}.
 
 P\'erez et al. showed that for both stellar and supermassive black holes in Schwarzschild and Kerr STVG spacetimes, the innermost stable circular orbit (ISCO) is always larger than for the corresponding spacetimes in GR; this occurs for all the values of the parameter $\alpha$ considered by the authors.  The major difference with respect to GR occurs for supermassive Kerr black holes ($a = 0.99$): the ISCO increases up to 716 percent with respect to the value of the ISCO in GR for $\alpha = 2.45$.
 
Notice that formulas (\ref{ene}) and (\ref{mom}) were derived from the laws of conservation of rest mass, angular momentum, and energy. These expressions, however, do not longer hold in STVG. A neutral test particle in STVG spacetime is subjected to a gravitational Lorentz force whose vector potential in Boyer-Lindquist coordinates is:
\begin{equation}
 \bm{\phi} = \frac{- Q r} {\rho^{2}}\left(\mathbf{dt} - a \sin{\theta}^{2} \mathbf{d\phi}\right).
\end{equation}
Here, $Q$ is the gravitational source charge of the vector field $\phi^{\mu}$. Expressions (\ref{ene}) and (\ref{mom}) are now redefined as:
\begin{eqnarray}
\tilde{E} & = & - \frac{p_{t}}{m} + \frac{q}{m}\phi_{t},\\
\tilde{L} & = & \frac{p_{\phi}}{m} + \frac{q}{m}\phi_{\phi},
\end{eqnarray}
and the conservation laws for angular momentum and energy take the form:
\begin{eqnarray}
\left(\tilde{L} - w\right)_{, r} + \frac{q}{m} \phi_{\phi_{,r}} & = & f \left(\tilde{L} + \frac{q}{m} \phi_{\phi}\right), \\ 
\left(\tilde{E} - \Omega w\right)_{, r} + \frac{q}{m}  \phi_{t_{,r}} & = & f \left(\tilde{E} + \frac{q}{m} \phi_{t }\right), 
\end{eqnarray}
 where
\begin{eqnarray}
 f & = & 4 \pi e^{\nu + \Phi + \mu} F / \dot{M_{0}}, \\
 w & = & 2 \pi e^{\nu + \Phi + \mu} {W_{\phi}}^{r} / \dot{M_{0}}.
\end{eqnarray}
$F$ denotes the emitted flux, ${W_{\phi}}^{r}$ the torque per unit circunference, $e^{\nu + \Phi + \mu}$ is the square root of the metric determinant, and $\dot{M_{0}}$ the mass accretion rate.

Once the heat $Q$ emitted by the disk is computed, the temperature profile can be obtained by means of the Stefan Boltzmann's law:\footnote{It is assumed that the disk is optically thick in the z-direction, so that every element of area on its surface radiates as a black body.}
\begin{equation}
 T(x) = z(x) \left(\frac{Q(x)}{\sigma}\right)^{1/4}. 
\end{equation}
$z(x)$ gives the correction due to gravitational redshift, and $\sigma$ denotes the Stefan-Boltzmann constant. Under the black body hypothesis, the emissivity per unit frequency  $I_{\nu}$ of each element of area of the disk is described by the Planck function:
\begin{equation}
I_{\nu}(\nu, x) = \frac{2 h \nu^{3}}{c^{2} \left[e^{\left(h \nu / k T(x)\right) } - 1\right]}.  
\end{equation}
Finally, the total luminosity at frequency $\nu$ is:
\begin{equation}
 L_{\nu}  =  \frac{4 \pi h G^{2} M^{2} \nu^{3}}{c^{6}}\int^{x_{out}}_{x_{isco}}  \frac{x \ dx}{\left[e^{\left(h \nu / k T(x)\right) }- 1\right]}.
\end{equation}

\begin{figure}[ht]
\centerline{
  \subfigure[Temperature as a function of the radial coordinate.]
     {\includegraphics[width=3in]{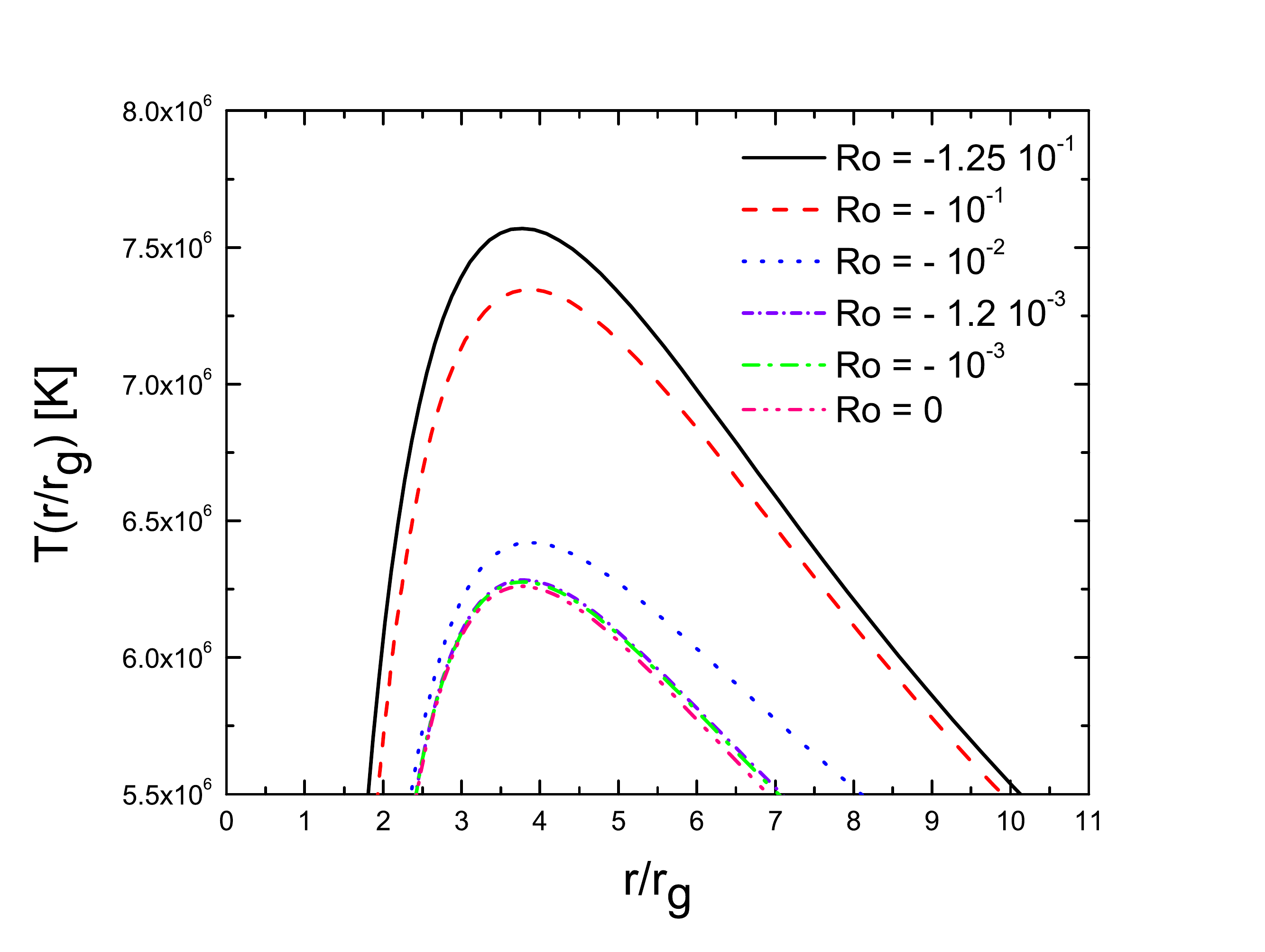}\label{fig1a}}
  \hspace*{4pt}
  \subfigure[Luminosity as a function of the energy.]
     {\includegraphics[width=3in]{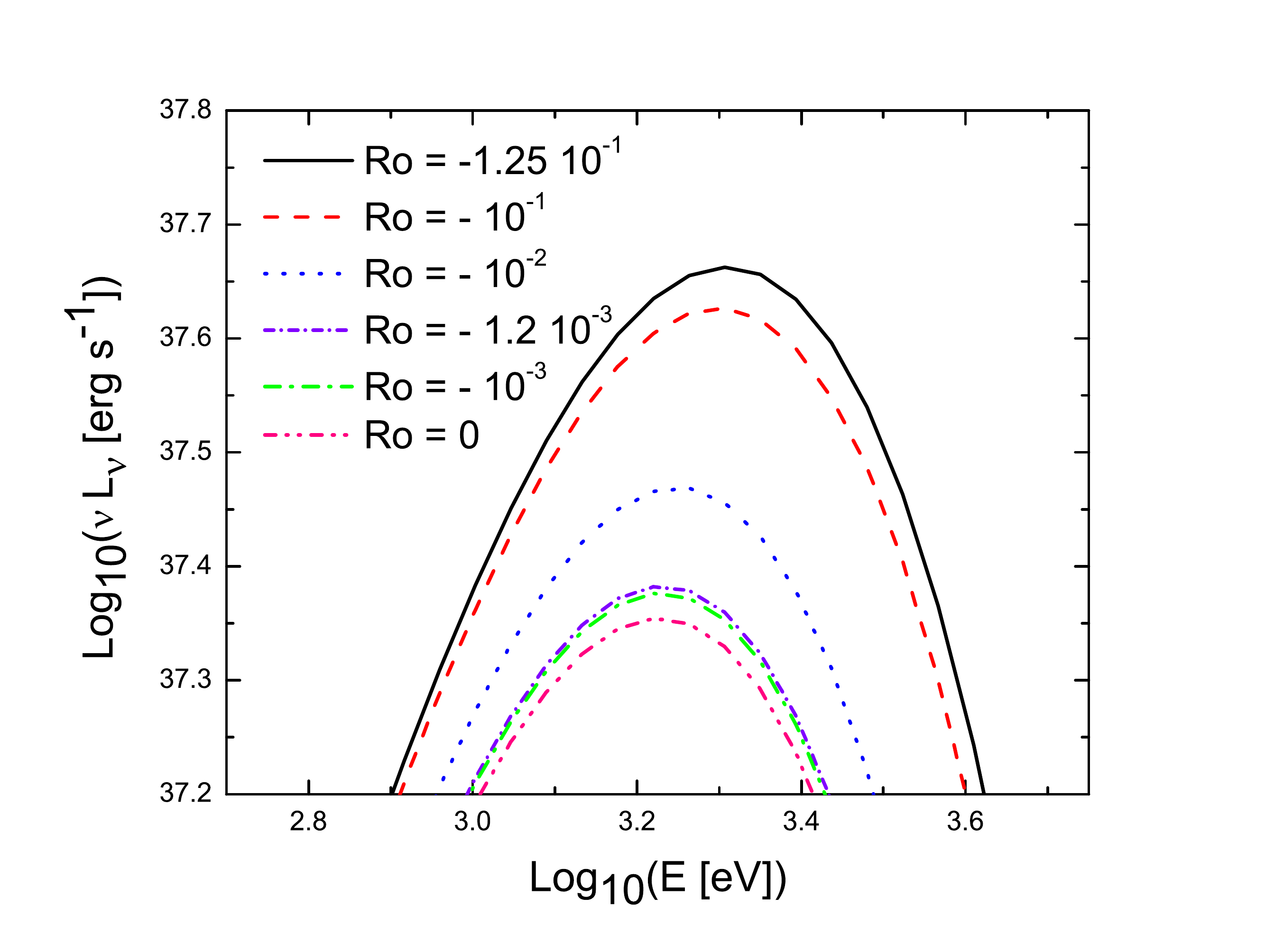}\label{fig1b}}
}
\caption{$f(R)$-Kerr black hole of angular momentum $\mathsf{a} = 0.99$ for $\mathsf{R_{0} < 0}$. From P\'erez et al.\cite{per+13}. Reproduced by permission of the authors.} \label{ra_fig2} 
\end{figure}

We show in Figures \ref{fig1a} and \ref{fig1b} the temperature and luminosity distributions for an accretion disk around a $f(R)$-Kerr black hole for negative values of the Ricci scalar\cite{per+13}. P\'erez and coworkers adopted for the values of the relevant parameters, i.e., $M$ mass, $\dot{M}$ accretion rate, and $a$ angular momentum of the black hole, the best estimates available for the galactic black hole Cygnus-X1\cite{gou+11,oro+11}. For $\mathsf{R_0} < 0$, the temperature and luminosity of the disk is always higher than in GR. In particular, for $\mathsf{R_0} = - 1.25 \times 10^{-1}$ the peak of the emission rises a factor 2 with respect to GR. In order to fit this $f(R)$-Kerr model  with current observations of Cygnus X-1 in the soft state, curvature values below $1.2 \times 10^{-3}$ have to be rule out. Accretion disk models for positive values of the Ricci scalar in the range $\mathsf{R_0} \in (0, 6.67 \times 10^{-4}]$ present no relevant differences compared to accretion disks in GR.

\begin{figure}[ht!]
\centerline{
  \subfigure[\textit{Top}: Temperature as a function of the radial coordinate. \textit{Bottom}: Residual plot of the temperature as a function of the radial coordinate.]
     {\includegraphics[angle=-90, width=3in]{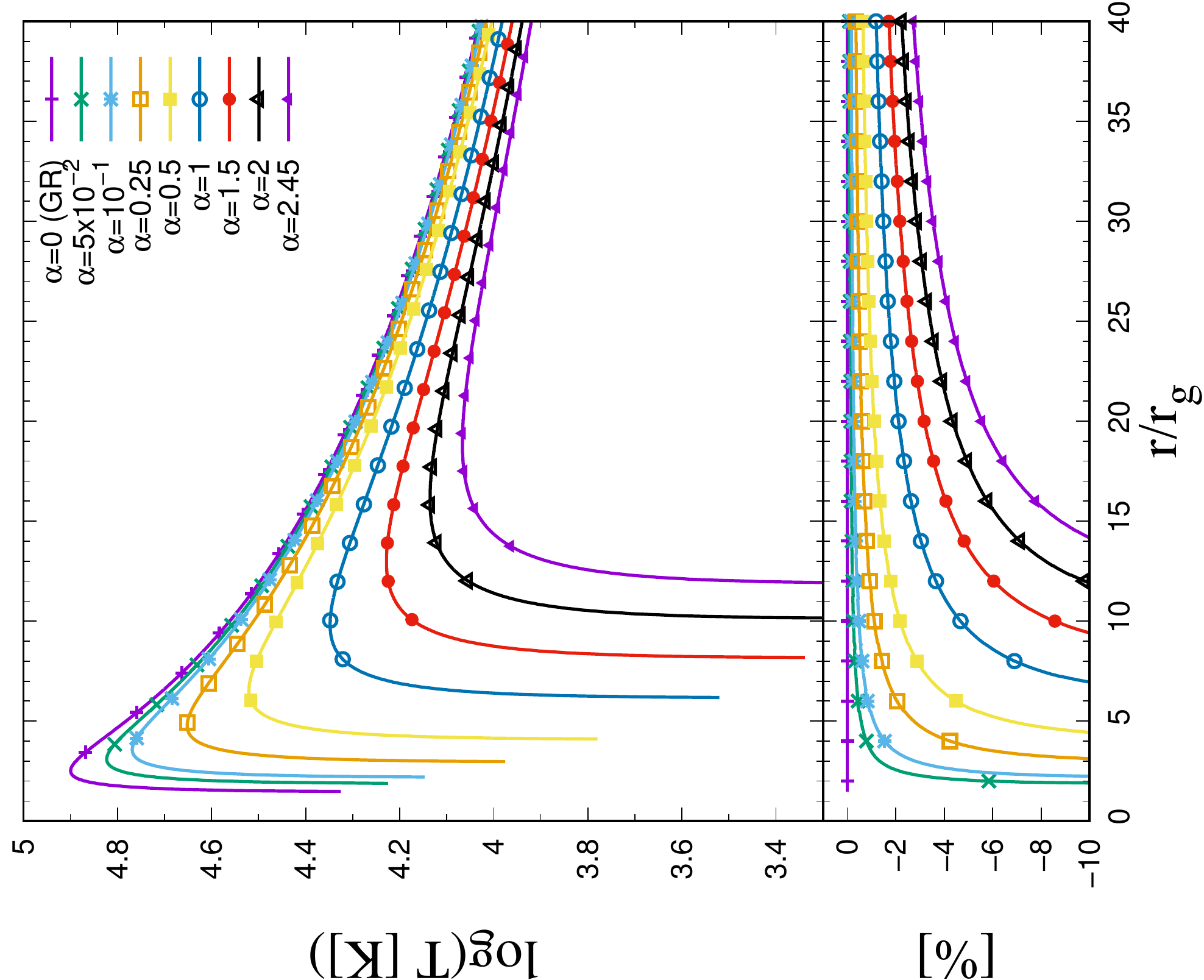}\label{fig2a}}
  \hspace*{4pt}
  \subfigure[\textit{Top}: Luminosity as a function of the energy. \textit{Bottom}: Residual plot of the luminosity as a function of the energy.]
     {\includegraphics[angle=-90,, width=3in]{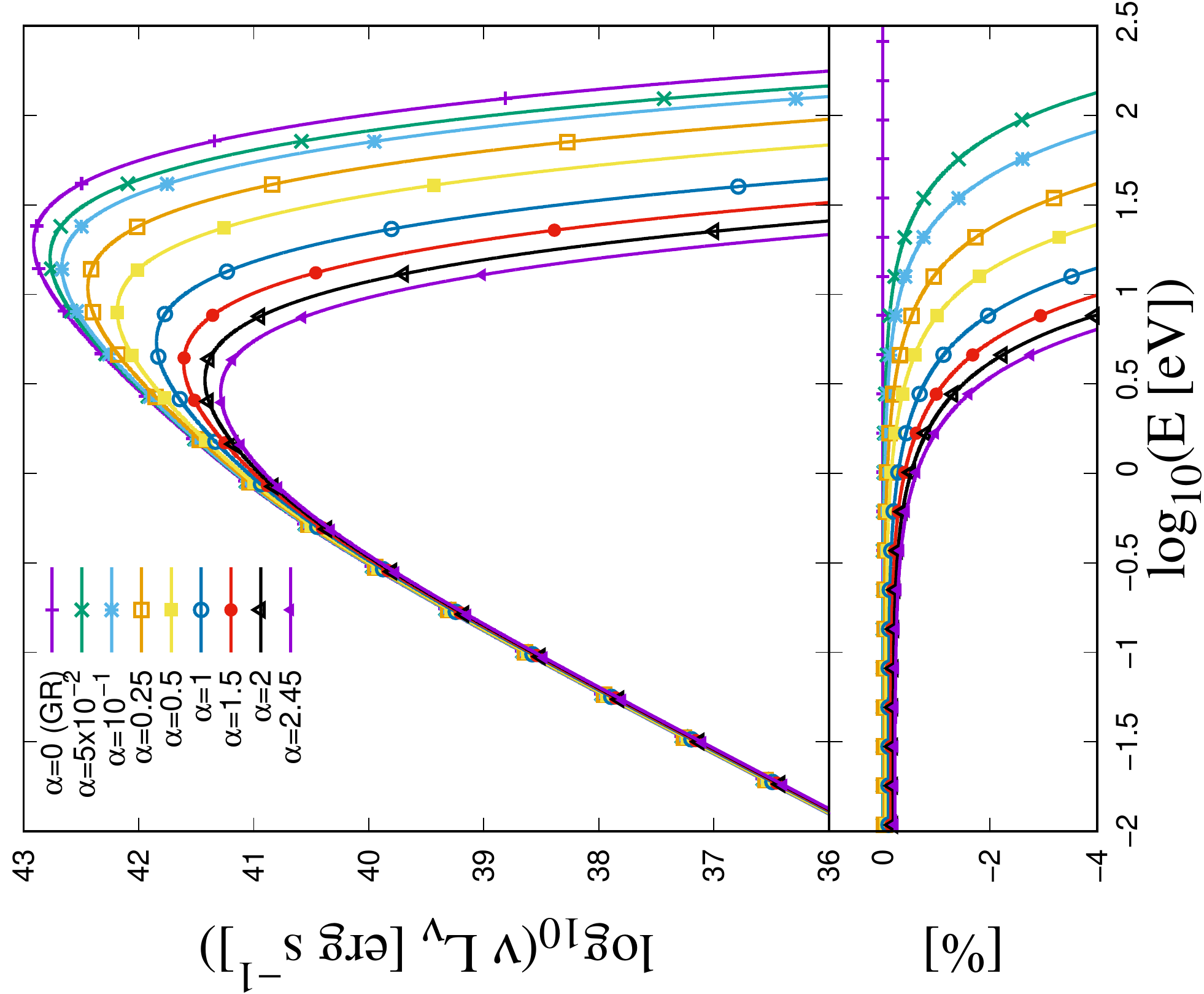}\label{fig2b}}
}
\caption{Supermassive Kerr-STVG black hole of spin $\tilde{a} = 0$. From P\'erez et al.\cite{per+17}. Reproduced by permission of the authors.} \label{ra_fig2} 
\end{figure}

According to P\'erez et al.\cite{per+17},  accretion disks around both Schwarzschild and Kerr black holes in STVG are colder and underluminuous in comparison with thin relativistic accretion disks in GR.  The greatest differences in temperature and luminosity were found for accretion disks around supermassive Kerr-STVG black holes, as despicted in Figures \ref{fig2a} and \ref{fig2b}. For instance, if $\alpha = 2.45$ the temperature decreases up to 12.30 percent, and the peak of the luminosity is about 4 percent lower than in GR. 

To sum up, we have seen that  investigations of accretion processes around black holes shed light on the behavior of gravity under extreme conditions. In particular, by comparing the spectral energy distributions predicted by modified theories of gravitation with current astronomical observations, more restrictive limits on the values of the free parameters of these theories can be set.  For $f(R)$-Schwarzschild and $f(R)$-Kerr black holes, thin stable accretion disks are possible for $\mathsf{R_{0}} \in (-\infty, 10^{-6}]$ and $\mathsf{R_{0}} \in  \left[- 1.2 \times 10^{-3}, 6.67 \times 10^{-4}\right]$, respectively.  In the case of Schwarzschild and Kerr STVG black holes,  accretion disks can exist for stellar sources if $0 < \alpha < 0.1$, while for supermassive sources $0.03 < \alpha < 2.47$.

It should be mentioned that other features of accretion flows in $f(R)$-gravity were studied by some authors. Ahmed and coworkers\cite{ahm+16,azr17} analyzed some aspects of the Michel-type accretion onto static spherically symmetric black holes in $f(R)$-gravity.  Effects of radial and angular pressure gradients on thick accretion disks in $f(R)$-Schwarzschild geometry with constant Ricci curvature were investigated by Alipour et al.\cite{ali+16}.

Bhattacharjee and collaborators\cite{bha+15} researched additional sources of vorticity in accreted material by black holes, others than the classical baroclinic instability. In particular, they analyzed how the spacetime geometry contributes to the vorticity generation in accretion disk plasma, both in GR and $f(R)$-gravity with constant Ricci scalar.  They found, as expected, that vorticity generation is more effective in Kerr than Schwarzschild spacetime, and also stronger in $f(R)$-gravity. The efficiency of this mechanism increases in the regions where strong gravity is dominant. Interestingly, the formalism developed by Bhattacharjee et al.\cite{bha+15} can be extended to multifluid species, and might provide a novel mechanism for angular momentum transport.


\section{Effects on jets}\label{Sec:jets}

The first observational evidence of the existence of astrophysical jets was due to Herber Curtis of Lick Observatory.  In 1918, he observed in M87, a supergiant elliptical galaxy in the constellation Virgo, ``a curious straight ray ... apparently connected with the nucleus  by a thin line of matter''. Nowadays, we know that such a ``straight ray'' is a collimated flow of particles and electromagnetic fields ejected by the supermassive black hole that lies at the core of the galaxy.  The highly collimated relativistic jet  extends at least 1.5 kiloparsecs from the nucleus of M87 well into the intergalactic medium.

Jets are observed in a plethora of astrophysical systems, from protostars to active galactic nuclei. There seems to be some key ingredients necessary for the formation of relativistic jets: accretion onto a spinning compact object, and the presence of a large-scale magnetic field.

It is though that the launching region of relativistic jets in active galactic nuclei is near the event horizon of the supermassive black hole, where the strong gravity effects are important. Thus, it is reasonable to expect that any deviations from GR should manifest on such scales. These effects on relativistic jets were recently investigated by Lopez Armengol and Romero\cite{lop+17a} in Scalar-Tensor-Vector Gravity. The aim of these authors was to compare GR and STVG close to the gravitational source and study the differences between the two theories on short scales\footnote{Up to the moment this chapter was written, no works on relativistic jets in $f(R)$-gravity were found.}.   

The method used by the authors was to calculate the trajectories of massive particles in Kerr-STVG spacetime. The black hole parameters $M$ mass, and $a$ angular momentum were adopted from the observational estimates  for the supermassive black hole in M87 reported by Gebhardt et al.\cite{geb+11}, and Li et al.\cite{li+09}. 

The results of Lopez Armengol and Romero  can be divided in two main parts. First, they studied the azimuthal orbits of massive particles set with initial position $r_{0} = 140 G_{\rm N} M/c^{2}$, $\theta_{0} = 0.18$, $\phi_{0} = 0$ and initial Lorentz factor $\gamma = 2$. The values of the parameters were taken from the observational results by Mertens et al.\cite{mer+16}. Two different cases for the ejection angles were taken into account: $\theta^{\rm A}_{\rm ej} = 0$ and $\theta^{\rm B}_{\rm ej} = 0.3$. 

The value of the free parameter $\alpha$ of the theory is related to the parameter $M_{0}$ by the formula:
\begin{equation}
  \alpha = \sqrt{\frac{M_{0}}{M}},
  \end{equation}
  where $M$ is the mass of the gravitational source. On the other hand, the parameter $\kappa$ that appears in the equation of motion for a test particle:
  \begin{equation}
 \left(\frac{d^{2} x^{\mu}}{d \tau^{2}} + \Gamma^{\mu}_{\alpha \beta} \frac{dx^{\alpha}}{d\tau} \frac{dx^{\beta}}{d\tau}\right) = \kappa {B^{\mu}}_{\nu} \frac{d x^{\nu}}{d \tau},   
  \end{equation}
  is also linked with the parameter $\alpha$ according to the following expression postulated by Moffat\cite{mof+15a}:
  \begin{equation}
   \kappa = \sqrt{\alpha G_{\rm N}}. 
  \end{equation}
  
 The calculation of the orbits was made for two runs of the parameters. In the first run, $M_{0}$ was fixed\footnote{Lopez Armengol and Romero\cite{lop+17a} established an upper limit for the value of $M_{0}$ using: 1) the observational estimates for the radius of $M87^{*}$ (supermassive black hole at the center of the elliptical galaxy M87) of $\approx 8 G_{\rm N} M/ c^{2}$ by Broderick et al.\cite{bro+15}, and 2) the formula for the event horizons of a Kerr-STVG black hole (see Section \ref{bh-stvg}, formula (\ref{hori-Kerr})).} ($M_{0} = 10^{11}$), and the parameter $\kappa$ was sample: $\kappa_{1} = 10^{2} \sqrt{\alpha G_{\rm N}}$,  $\kappa_{2} = 10^{3} \sqrt{\alpha G_{\rm N}}$, $\kappa_{3} = 10^{4} \sqrt{\alpha G_{\rm N}}$. Significant deviations from GR were found: the angular velocity $\omega_{\phi}$ as a function of the $z$ coordinate is enhanced or diminished by the gravito-magnetic forces depending on the initial ejection angle. Also, the augmented gravito-electric repulsion due to the growing values of $\kappa$ increases the local Lorentz factor $\gamma$ of the particles with time. The later implies that in the strong field regime of STVG, particles gravitationally
accelerate. 

The values of the local Lorentz factor $\gamma$, however, cannot be arbitrarily large.  The highest value of $\gamma$ observationally estimated for M87 corresponds to the spine of jet and is $\gamma \approx 10$\cite{mer+16}. Thus, the upper limit imposed to $\kappa$ is\cite{lop+17a}:
\begin{equation}
 \kappa \leq  10^{2} \sqrt{\alpha \ G_{\rm N}}.
\end{equation}

In the second run, Lopez Armengol and Romero took Moffat's weak field limit prescription $\kappa = \sqrt{\alpha G_{\rm N}}$, and set  $\theta^{\rm A}_{\rm ej} = 0$. In this case, the parameter $M_{0}$ was modified: $M_{0} = 10^{10} M_{\odot}$, $M_{0} = 10^{11} M_{\odot}$, and $M_{0} = 10^{12} M_{\odot}$.  Though the latter value violates the condition $M_{0} \leq 10^{11} M_{\odot}$ computed by\cite{lop+17a} taking into account observational constraints for $M87^{*}$, it was included for consistency checks.

Contrary to the first run, the effects of the Lorentz-like forces on trajectories became insignificant. This could be due to the small value of $\kappa$. The increment in the value of $\alpha$, through the parameter $M_{0}$, led to a larger decrement of the particle velocity in contrast to GR. 

\begin{figure}[hb!]
\centerline{
  \subfigure[Projection in the  $x$-$z$ plane of the trajectories of massive particles at the base of jet, for different values of $\kappa$. The filled region represents the jet as modeled by Mertens et al.\cite{mer+16}.]
     {\includegraphics[angle=-90, width=3in]{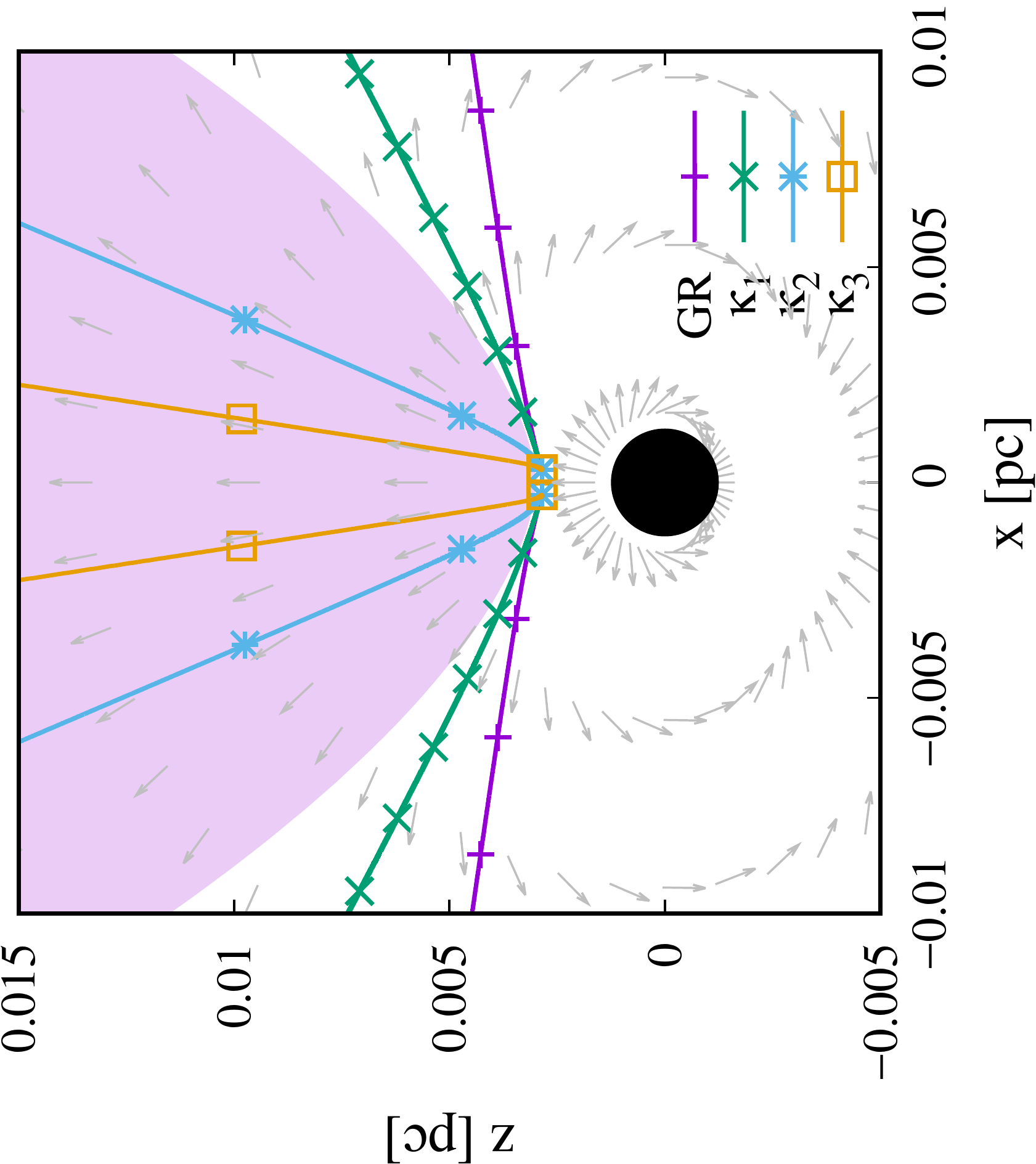}\label{fig3a}}
  \hspace*{4pt}
  \subfigure[Angular velocity as a function of the $z$ coordinate for particles ejected with a wide angle at the launching region of the jet.]
     {\includegraphics[angle=-90,width=3in]{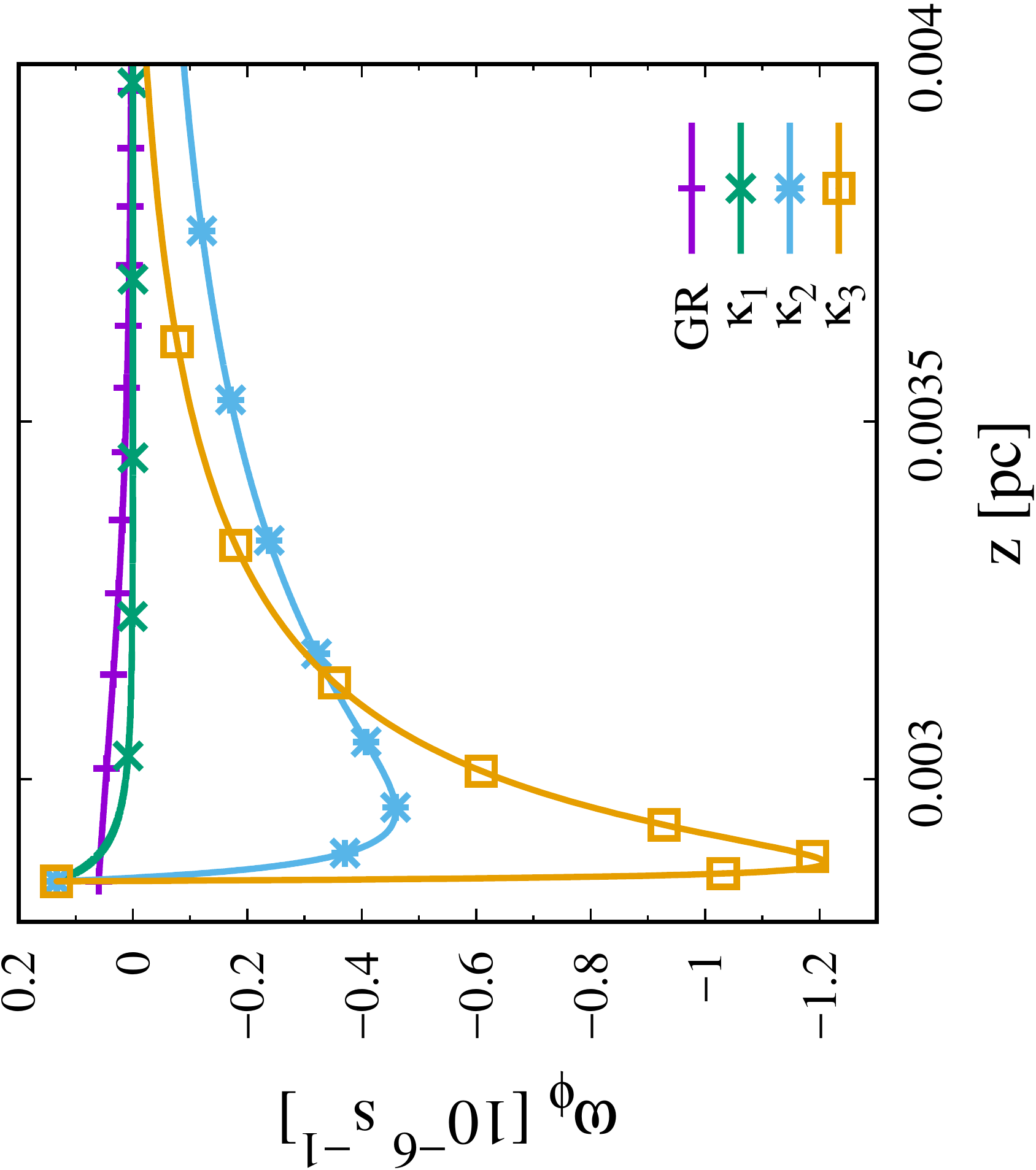}\label{fig3b}}
}
\caption{The gravito-magnetic field in STVG severely changes the trajectories of particles giving rise to effects that are absent in GR. Reprinted by permission from Springer Customer Service Centre GmbH: Springer Nature, Astrophysics and Space Science, Lopez Armengol, F.G. Romero, G.E. Astrophys Space Sci (2017) 362: 214. https://doi.org/10.1007/s10509-017-3197-6. } \label{ra_fig3} 
\end{figure}

Finally, the authors compared observational results on the formation zone of the jet in M87 with predictions of STVG. The values selected to model the jet were $r_{0} = 5 R_{\rm S}$, $M_{0} = 10^{10} M_{\odot}$, $\kappa_{1} = 10 \ \sqrt{\alpha G_{\rm N}}$, $\kappa_{2} = 10^{2} \ \sqrt{\alpha G_{\rm N}}$, $\kappa_{3} = 10^{3} \ \sqrt{\alpha G_{\rm N}}$; a wide ejection angle  was assumed in accordance to the Blandford-Payne mechanism for jet launching\cite{bla+82, spr10}. In Figure \ref{fig3a}, the $x$-$z$ trajectories for different values of $\kappa$ are plotted. The filled region corresponds to the jet according to the parametrization by Mertens et al.\cite{mer+16}. Notably, the larger the values of $\kappa$, the higher is the collimation of the jet.

The plot of the angular velocity $\omega_{\phi}$ as a function of $z$ for different values of $\kappa$ is shown in Figure \ref{fig3b}. At the base of the jet, the gravito-magnetic force leads to a counter rotation in $\phi$. Since the field lines rotates along the trajectory, from certain value of $z$, the sign of the gravito-magnetic force changes and $\omega_{\phi}$ starts to increase. The latter result is particularly interesting in the light of the recent observations of the jet of M87\cite{mer+16} in which, for the first time, rotation has been directly measured. The jet component closer to the launching region rotates clockwise while further from M87, the jet rotates counterclockwise\cite{mer+16}. The calculations of Lopez Armengol and Romero for the scale where rotation changes sign and the order of magnitude of the angular velocity $\omega_{\phi}$ are in agreement with the estimations by Mertens and coworkers\cite{mer+16}.

In conclusion, we have seen that STVG offers an alternative mechanism for jet formation of purely gravitational origin that is compatible with current observational data.


\section{Gravitational waves}

In this section we will focus on the implications for $f(R)$-gravity and Scalar-Tensor-Vector Gravity of the recent detection of gravitational waves by the the LIGO Scientific Collaboration and Virgo Collaboration.\footnote{The reader interested in more theoretical aspect of gravitational waves in $f(R)$-gravity is referred to\cite{sot+10,cap+11}.}

Capozziello and collaborators\cite{cap+08} were the first to generalize some previous results on gravitational waves in $f(R)$-gravity. They showed that there are only three propagating  degrees of freedom in such theory: the massless plus and cross polarizations which are the same as in GR, and a third massive scalar mode given by a mixed of longitudinal and transverse polarizations (see also Liang et al.\cite{lia+17}).

The signal that a gravitational wave detector would identify if gravitational waves had such additional polarization modes were computed by Corda\cite{cor18}. He provided expressions for the frequency and angular dependent response function of a gravitational wave interferometric detector in the presence of a third massive mode (see Eq. 59 in the paper by Corda \cite{cor18}). The formulas given by Corda are of particular importance since they allow to discriminate between massless and massive modes in Scalar Tensor Gravity and $f(R)$-gravity.

Current ground based gravitational wave interferometers do not have the sensitivity to  detect the directions in which the instrument is oscillating. For the transient waves already observed, a study of the gravitational wave polarization cannot yet be done with the LIGO-Virgo network, since at least five non-co-oriented  arm antenas are required to break the degeneracies between the five distinguishable modes of a generic metric theory of gravity\cite{cor18, abb+18}.  Nonetheless, studies of the polarization content of the signal have already been conducted. This was the case for the gravitational wave event GW170814 produced by the merger of two stellar mass black holes observed with both the LIGO and Virgo detectors\cite{abb+17a1}; the analysis of the data strongly favor pure tensor polarization of gravitational waves, over pure scalar or pure vector polarizations.

Theory independent polarizations measurements, however, can be done with the current instruments if the detectors are exposed to sufficiently long signals. This occurs for pulsars which are expected to emit continuous gravitational waves. At the beginning of 2018, the LIGO Scientific Collaboration and Virgo Collaboration\cite{abb+18} presented the results of the first directed search of nontensorial gravitational waves. The investigation focused on 200 known pulsars using data from aLIGO's first observation run; no assumption was made about the polarization modes of the gravitational waves. The data showed no evidence for the emission of gravitational signals of tensorial or nontensorial polarization from any of the pulsars studied. They also obtained upper limits for the strain of the scalar and vector modes ($1.5 \times 10^{-26}$ at 95 percent credibility) that can in turn be used to constraint alternative theories of gravity. Notice that an important assumption of the work was that the gravitational wave emission frequency $f_{\rm GW}$ was twice of the rotational frequency $f_{\rm rot}$ of the source, that is $f_{\rm GW} = 2 f_{\rm rot}$. This hypothesis follows the most favored emission model in GR. Such condition will be relaxed in future investigations.  

The Advance LIGO-Virgo network has also carried on a search for a generically-polarized stochastic background of gravitational waves\cite{abb+18a} using data from Advanced
LIGO’s O1 observing run.\footnote{The gravitational wave background is a random gravitational wave signal that is supposed to be generated by distant compact binary mergers, core-collapse supernovae, and rapidly-rotating neutron stars; a background of cosmological origin may also be present (see for instance\cite{cap+18}).} No evidence was found for
the presence of a background of gravitational waves, and of any polarization. However, direct upper limits  were established on the contributions of vector and scalar modes to the stochastic background.

A key point recently clarified by Corda\cite{cor18} is that the constraints on the graviton mass derived by the LIGO Scientific Collaboration and the Virgo Collaboration\cite{abb+17a} are not on some extra polarization mode of the gravitational waves but on the tensor modes.  In fact, the upper limit on the graviton mass is derived assuming that gravitons disperse in vacuum as massive particles; the lack of dispersion in the gravitational waves sets a lower limit on the  Compton wavelength $\lambda_{\rm g} > 1.6 \times 10^{13}$ km, or for the graviton mass $m_{\rm g} < 7.7. \times 10^{-23} \ {\rm eV/c^{2}}$\cite{abb+17a}.

Some authors have used the constraint on the graviton mass mentioned above to bound free parameters in different models of $f(R)$-gravity. For instance, Vainio and Vilja\cite{vai+17} constrained the parameter $\mu$ in the Hu-Sawicki model\cite{hu+07} (taking $n \approx 1$), and also the parameter $\lambda$ in Starobinsky model\cite{sta07}. A similar procedure was employed by Lee\cite{lee17b} to obtain limits on the parameter $M_{0}$ of a general constructed $f(R)$-model determined from cosmological observations\cite{lee17a}. 

On August 17th 2017, it was observed for the first time a neutron star merger in gravitational waves\cite{abb+17b} (GW 170817). The electromagnetic counterpart\cite{abb+17c,abb+17d}, GRB 170817A, was detected 1.7 s after GW170817. The observed delay between both events was used to restrict the difference between the speed of gravitational waves and the speed of light:
\begin{equation}
\abs{\frac{c_{\rm GW}}{c}-1} < 5 \times 10^{-16},
\end{equation}
where $c_{\rm GW}$ denotes the speed of the gravitational waves, and $c$ stands for the speed of light. This result was used by some authors to show that a large class of alternative theories of gravitation are on the verge of being completely discarded (see, for instance, Refs.\cite{ezq+17, par+18}). Some theories, however, have survived such stringent test. Nojiri and Odintsov\cite{noj+18} explicitly showed that in $f(R)$-gravity the propagating light speed $c$ is identical to the propagating speed $c_{\rm GW}$ of the gravitational waves. There is some difference, however, in the gravitational wave propagation phase with respect to the light one; the authors suggest future observations that could detect the shift of the phase. 

A novel generalized framework for testing the nature of gravity using gravitational wave propagation has just been introduced by Nishizawa\cite{ni17}. The method employed was to analytically solve the gravitational wave propagation equation in an effective field theory for dark energy and get a WKB solution. The gravitational waveform obtained contains functions of time that characterize modified amplitude damping, modified propagation speed, non zero graviton mass, and also a source term for the gravitational waves. The author also provided specific expressions of these general functions in the context of various alternative theories of gravitation. In a second paper, Arai and Nishizawa\cite{ara+17} applied this generalized framework to test all possible models of the Horndeski theory. Using the data from the simultaneous detection of GW170817 and GRB170817A, some models within the theory were excluded while quintessence, nonlinear kinematic theory, and $f(R)$-gravity were favored.

What was the fate of Scalar Tensor Vector Gravity after the six gravitational wave detections by the LIGO-Virgo Collaboration? 

As we already mentioned, the detection of gravitational waves produced by the merger of a binary system of neutron stars, and the subsequent observation of a short Gamma-ray Burst (GRB170817A), provided invaluable data to test GR and also alternative theories of gravitation.

A classical test of GR is the Shapiro time delay, or gravitational time delay\cite{sha64}. GR predicts that the amount of time it takes an electromagnetic signal to travel to a  target and return is longer if a massive object is close in its path. The time delay is caused by spacetime dilatation, which increases the path length. In GR, the Shapiro time delay is the same for gravitational waves and photons since both travel on null geodesics. This may not be the case in other theories of gravity, in particular in those modified theories of gravity which dispense of dark matter and reproduce the modified newtonian dynamics (for instance galaxy rotation curves) in the non-relativistic limit, the so-called \textit{Dark Matter Emulators}\cite{kah+07}.

Theories of gravity dubbed ``Dark Matter Emulators'' were defined explicitly as\cite{kah+07,bor+18}:
\begin{enumerate}
\item Ordinary matter couples to the metric $\tilde{g_{\mu \nu}}$ ($\tilde{g}$ denotes the “disformally transformed metric”) that would be produced by GR with dark matter, and\\
\item  Gravitational waves couple to the metric $g_{\mu \nu}$ produced by GR without
dark matter.
\end{enumerate}

From the above definitions it follows that in Dark Matter Emulators, photons suffer an additional Shapiro time delay due to the dark matter needed if GR was correct. Boran and collaborators\cite{bor+18} estimated that the Shapiro time delay due to the gravitational potential of the total dark matter distribution along the line of sight from NGC 4993 to the Earth is of the order of 400 days.  Since the electromagnetic detection of GW170817 was almost immediate while Dark Matter Emulators predict delays over a year, Boran et al. concluded that these theories were rule out. 

In the realm of Dark Matter Emulators, Boran et al. included Scalar-Tensor-Vector Gravity, which we have extensively discussed in this chapter. Green, Moffat and Toth\cite{gre+18} have recently demonstrated that STVG do not belong to such class of theories.

Green et al.\cite{gre+18} explicitly showed that in STVG:
\begin{itemize}
\item Gravitational waves move at the speed of light as photons do.\\

STVG is constructed on these gravitational fields: the metric $g_{\mu \nu}$ (a spin 2 massless graviton), a scalar field $G = G_{\rm N} \left( 1 + \alpha \right)$ (a spin 0 massless graviton), a vector field $\phi_{\mu}$ (a spin 1  repulsive massive graviton), and  a spin 0 scalar field denoted $\mu$ that  is the mass of the vector field $\phi_{\mu}$. The mass of $\mu$, represented by $m_{\phi}$ was estimated from fits to galaxy rotation curves and clusters without dark matter\cite{mof+13,mof+14}, and it is $m_{\phi} = 2.8 \times 10^{-28}$ eV, which is of the order of the experimental bound on the photon mass. Since the vector particle mass is extremely small, the three gravitons that corresponds to $g_{\mu \nu}$, $G = G_{\rm N} \left( 1 + \alpha \right)$, and $\phi_{\mu}$ move at the speed of light.\\

\item The Weak Equivalence Principle is satisfied.\\

The equation of motion of a massive test particle is STVG is\cite{mof06}:
\begin{equation}
m \  \left(\frac{d^{2} x^{\mu}}{d \tau^{2}} + \Gamma^{\mu}_{\alpha \beta} \frac{dx^{\alpha}}{d\tau} \frac{dx^{\beta}}{d\tau}\right) = q_{g} {B^{\mu}}_{\nu} \frac{d x^{\nu}}{d \tau},   
  \end{equation}
where $u^{\mu} = dx^{\mu}/ds$ and $q_{g} = \kappa m$ is the gravitational charge of the test particle\cite{mof06}  (see also Section \ref{Sec:jets}). We can cancel the mass on both sides of the equation above thus showing that all particles (bodies) move on spacetime independently of their constitution.
  \end{itemize}

We see, then, that STVG is still standing after the observations of GW170817 and  GW170817A. Shall it be the case in the future? We do not know. What we do know is that further research is needed on the nature of gravitational waves in STVG. To date,  it has been only be derived the linearized field equations and gravitational energy flux in the weak field regime\cite{mof16}. In the case of binary pulsar well-separated the tensor radiated power reduces to the corresponding result in GR\cite{mof16} in agreement with observations of binary pulsars\footnote{Moffat obtained in STVG constraints of some parameters related to the spins of spiralling binary black holes. He also showed that STVG predicts a misaligning of the black holes that are in coalesce\cite{mof17}.}. 

The process of the merger of a system of binary black holes can be described in three phases: the inspiral, merger, and ringdown. The final stage, the ringdown, is when the resulting black hole horizon settles down through damped oscillations which can in turn be characterized by the Quasi-Normal Modes (QNMs).

In the first gravitational wave event detected\cite{abb+16a}, GW150914, it was possible to obtain the frequency and damping time of the QNMs. This data was used to derive the mass $M$ and spin $a$ of the black hole\cite{abb+16b}. If in the future additional QNMs are detected, i.e., first overtone, it may be possible to distinguish deviations from GR. For instance, the frequency of QNMs of Schwarzschild black holes in STVG is greater than in GR as the value of the free parameter $\alpha$ of the theory increases\cite{man+18} (see also Wei et al.\cite{wei+18} for some related results for Kerr black holes in STVG). In $f(R)$-gravity, Bhattacharyya and Shankaranarayanan\cite{bha+17} showed that vector and scalar perturbations in $f(R)$ black hole spacetimes do not emit the same amount of gravitational wave energy. Indeed, one of the most relevant differences between GR and $f(R)$-gravity is the existence, in the latter, of dipole and monopole radiation. This is a consequence of the additional degree of freedom of the theory represented by the scalar field\footnote{The computation of the gravitation radiation has been done for different choices of the $f(R)$ function; see for instance\cite{naf+11, ber+11, del+11}. }.


\section{Singularities and beyond}

Singularities appear under several circumstances in GR. Gravitational collapse, in quite general conditions, results in singularities \cite{Penrose1965}. Cosmological models of the expanding universe are also singular for normal matter contents and GR \cite{Hawking-Penrose1970}. All these singularities are pathological features, since they imply a breakdown of the corresponding models and their predictive power \cite{rom+2013}. Singularities are not entities of any kind, but a sign of the incompleteness of the underlying theory. They are not expected to show up in a consistent theory of quantum gravity.  But also classical theories can be free of singularities if they assume a more complex structure for spacetime.  

Non-singular black holes have been discussed in the framework of $f(R)$-theories by Olmo and Rubiera-Garcia \cite{Olm+15b,Olm+16}. They have found that in quadratic gravity and for anisotropic fluids a region with non-trivial topology (a so-called wormhole) replaces the singularity.  The resulting spacetime is geodesically complete although curvature divergencies exists in the wormhole throat.  With more generality Bejarano et al. \cite{Bej+17} have investigated the relation between energy density, curvature invariants, and geodesic completeness in such spacetimes. They have explored how the anisotropic fluid helps to modify the innermost geometry of the black hole.  A variety of configurations with and without wormholes have been found, as well as solutions with de Sitter interiors, solutions that mimic non-linear models of electrodynamics coupled to GR, and configurations
with up to four horizons. The fact that several of these models show divergences in the curvature scalar but nevertheless remain geodesically complete shows that simplistic analyses based only on the behaviour of curvature invariants can be misleading. A remarkable feature is that the anisotropic fluid has a stress-energy tensor satisfying the energy conditions, which are usually violated to generate regular black holes in GR (see, for instance, the work of P\'erez et al. \cite{Per+14}).

Regular black holes can be also obtained in STVG gravity. Moffat has discussed both static and rotating solutions \cite{mof+15a,mof+18}. In the case of the Scharzschild-like black hole the solution reads:
\begin{eqnarray}
  ds^2&=&\biggl(1-\frac{2GMr^2}{(r^2+\alpha G_NGM^2)^{3/2}}+\frac{\alpha G_NGM^2r^2}{(r^2+\alpha G_NGM^2)^2}\biggr)dt^2\nonumber\\
&-&\biggl(1-\frac{2GMr^2}{(r^2+\alpha G_NGM^2)^{3/2}}+\frac{\alpha G_NGM^2r^2}{(r^2+\alpha G_NGM^2)^2}\biggr)^{-1}dr^2\nonumber\\
& - & r^2d\Omega^2.
\label{MoffatBH}
\end{eqnarray}
This metric is regular at $r=0$ and asymptotically flat as $r\rightarrow\infty$.   For large $r$ it reproduces the Schwarzschild metric.  For small $r$ the metric behaves as
\begin{equation}
\label{deSitterMog}
ds^2=(1-\frac{1}{3}\Lambda r^2)dt^2-(1-\frac{1}{3}\Lambda r^2)^{-1}dr^2-r^2d\Omega^2,
\end{equation}
where the effective cosmological constant $\Lambda$ is given by
\begin{equation}
\Lambda=\frac{3}{G_N^2M^2}\biggl(\frac{\alpha^{1/2}-2}{\alpha^{3/2}(1+\alpha)}\biggr).
\end{equation}
So the interior material of the regular STVG  black hole satisfies the vacuum equation of state $p=-\rho$, where $p$ and $\rho=\rho_{\rm vac}$ are the pressure and the vacuum density, respectively. The effective cosmological constant $\Lambda$ can be positive or negative depending on the magnitude of $\alpha$, so that the interior of the black hole is described by either a de Sitter or anti-de Sitter spacetime. Shadows of these black holes are expected to be different from GR black holes of the same mass. Hence, Event Horizon Telescope observations can be used to test the theory \cite{mof+15a}.

Another interesting feature of STVG is that it allows for wormhole solutions that do not require exotic matter. If wormholes exist in nature they might be found by gravitational lensing, as pointed out by Safonova et al.  \cite{Saf02}. Wormhole lensing events and wormhole shadows are expected to differ in STVG with respect to GR. 

One interesting feature of $f(R)$-gravity when applied to the whole universe is that early time cosmology can be adjusted in such a way as to guarantee the existence of homogeneous and isotropic models that avoid the Big Bang singularity. In such models the Big Bang singularity can be replaced by a cosmic bounce without violating any energy condition. The bounce is possible even for pressureless dust \cite{Barr+09,Paul+14,Bat+16}. Such models can be tested through measurements of the modes B of the CMB polarization by instruments as the forthcoming QUBIC \cite{QUBIC+16}. Specifically, it has been shown that there are distinctive (oscillatory) signals on the primordial gravitational wave spectrum for very low frequencies in $f(R)$-gravity; such signals correspond to modes that are currently entering the horizon \cite{Bou+13}.

Detailed studies regarding the potential role of STVG in producing cosmological bounces are still missing, likely because of the complexity of the theory. Recently, Jamali et al. \cite{Jam+18} have shown that extra fields in STVG cannot provide a late time accelerated expansion. Furthermore, they have solved the non-linear field equations numerically and calculated the angular size of the sound horizon which results outside the current observational bounds. However, further research is necessary since the models analyzed so far are rather simple, without self-interactions. 

Just as in the last century the universe was investigated through the whole electromagnetic spectrum, the 21st century is the starting of a new time in which gravitational waves  will be detected over a wide range of frequencies. A new generation of gravitational waves detectors in space and underground, eLISA, KAGRA, and the Einstein Telescope are planned to start working in the next decades. Pulsar timing monitoring will also make possible the investigation of the exceedingly long wavelength perturbations in spacetime curvature caused by the merging of supermassive black holes\cite{Arz+18,Ell+14}. Alternative theories of gravity, and GR itself, will be put to the test to a level of detail that is impossible today. Novel aspects of the nature of gravity might be unveiled that perhaps none of the actual theories predict or foresee.

\section*{Acknowledgments}

We thank S.E. Perez Bergliaffa, John Moffat, F. L\'opez Armengol, and L. Combi for many discussions on these topics. This work was supported by CONICET (PIP2014-0338) and
by the Spanish Ministerio de Economía y Competitividad (MINECO) under grant AYA2013-47447-C3-1-P and AYA2016-76012-C3-1-P.


\bibliographystyle{ws-rv-van}
\bibliography{bibliography}

\begin{thebibliography}{171}
\providecommand{\natexlab}[1]{#1}
\providecommand{\url}[1]{\texttt{#1}}
\expandafter\ifx\csname urlstyle\endcsname\relax
  \providecommand{\doi}[1]{doi: #1}\else
  \providecommand{\doi}{doi: \begingroup \urlstyle{rm}\Url}\fi

\bibitem{Einstein1916}
A.~{Einstein}, {Die Grundlage der allgemeinen Relativit{\"a}tstheorie},
  \emph{Annalen der Physik}. {\bf 354}, \penalty0 769--822  (1916).
\newblock \doi{10.1002/andp.19163540702}.

\bibitem{Will2014}
C.~M. {Will}, {The Confrontation between General Relativity and Experiment},
  \emph{Living Reviews in Relativity}. 17:\penalty0 4  (June, 2014).
\newblock \doi{10.12942/lrr-2014-4}.

\bibitem{abb+16a}
B.~P. {Abbott}, R.~{Abbott}, T.~D. {Abbott}, M.~R. {Abernathy}, F.~{Acernese},
  K.~{Ackley}, C.~{Adams}, T.~{Adams}, P.~{Addesso}, R.~X. {Adhikari}, and
  et~al., {Observation of Gravitational Waves from a Binary Black Hole Merger},
  \emph{Phys. Rev. Lett.} 116\penalty0 (6):\penalty0 061102  (Feb., 2016).
\newblock \doi{10.1103/PhysRevLett.116.061102}.

\bibitem{Penrose1965}
R.~{Penrose}, {Gravitational Collapse and Space-Time Singularities},
  \emph{Phys. Rev. Lett.} {\bf 14}, \penalty0 57--59  (Jan., 1965).
\newblock \doi{10.1103/PhysRevLett.14.57}.

\bibitem{Hawking-Penrose1970}
S.~W. {Hawking} and R.~{Penrose}, {The Singularities of Gravitational Collapse
  and Cosmology}, \emph{Proceedings of the Royal Society of London Series A}.
  {\bf 314}, \penalty0 529--548  (Jan., 1970).
\newblock \doi{10.1098/rspa.1970.0021}.

\bibitem{Giddings1995}
S.~B. {Giddings}, {The black hole information paradox}, \emph{ArXiv High Energy
  Physics - Theory e-prints}  (Aug., 1995).

\bibitem{cap+11}
S.~{Capozziello} and V.~{Faraoni}, \emph{{Beyond Einstein Gravity}}  (2011).
\newblock \doi{10.1007/978-94-007-0165-6}.

\bibitem{mof06}
J.~W. {Moffat}, {Scalar tensor vector gravity theory}, \emph{J. Cosm.
  Astropart. Phys.} 3:\penalty0 004  (Mar., 2006).
\newblock \doi{10.1088/1475-7516/2006/03/004}.

\bibitem{baa+34}
W.~{Baade} and F.~{Zwicky}, {On Super-novae}, \emph{Proceedings of the National
  Academy of Science}. {\bf 20}, \penalty0 254--259  (May, 1934).
\newblock \doi{10.1073/pnas.20.5.254}.

\bibitem{opp+39}
J.~R. {Oppenheimer} and G.~M. {Volkoff}, {On Massive Neutron Cores},
  \emph{Phys. Rev.} {\bf 55}, \penalty0 374--381  (Feb., 1939).
\newblock \doi{10.1103/PhysRev.55.374}.

\bibitem{tol39}
R.~C. {Tolman}, {Static Solutions of Einstein's Field Equations for Spheres of
  Fluid}, \emph{Phys. Rev.} {\bf 55}, \penalty0 364--373  (Feb., 1939).
\newblock \doi{10.1103/PhysRev.55.364}.

\bibitem{hew+67}
A.~{Hewish}, S.~J. {Bell}, J.~D.~H. {Pilkington}, P.~F. {Scott}, and R.~A.
  {Collins}, {Observation of a Rapidly Pulsating Radio Source}, \emph{Nature}.
  {\bf 217}, \penalty0 709--713  (Feb., 1968).
\newblock \doi{10.1038/217709a0}.

\bibitem{shkl67}
I.~S. {Shklovsky}, {On the Nature of the Source of X-Ray Emission of SCO
  XR-1.}, \emph{Astrophys. J. Lett.} {\bf 148}, \penalty0 L1  (Apr., 1967).
\newblock \doi{10.1086/180001}.

\bibitem{gia+62}
R.~{Giacconi}, H.~{Gursky}, F.~R. {Paolini}, and B.~B. {Rossi}, {Evidence for x
  Rays From Sources Outside the Solar System}, \emph{Phys. Rev. Lett.} {\bf 9},
  \penalty0 439--443  (Dec., 1962).
\newblock \doi{10.1103/PhysRevLett.9.439}.

\bibitem{san+66}
A.~{Sandage}, P.~{Osmer}, R.~{Giacconi}, P.~{Gorenstein}, H.~{Gursky},
  J.~{Waters}, H.~{Bradt}, G.~{Garmire}, B.~V. {Sreekantan}, M.~{Oda},
  K.~{Osawa}, and J.~{Jugaku}, {On the optical identification of SCO X-1},
  \emph{Astrophys. J.} {\bf 146}, \penalty0 316  (Oct., 1966).
\newblock \doi{10.1086/148892}.

\bibitem{cam07}
M.~{Camenzind}, \emph{{Compact objects in astrophysics : white dwarfs, neutron
  stars, and black holes}}. {Springer-Verlag}  (2007).
\newblock \doi{10.1007/978-3-540-49912-1}.

\bibitem{abb+17b}
B.~P. {Abbott}, R.~{Abbott}, T.~D. {Abbott}, F.~{Acernese}, K.~{Ackley},
  C.~{Adams}, T.~{Adams}, P.~{Addesso}, R.~X. {Adhikari}, V.~B. {Adya}, and
  et~al., {GW170817: Observation of Gravitational Waves from a Binary Neutron
  Star Inspiral}, \emph{Phys. Rev. Lett.} 119\penalty0 (16):\penalty0 161101
  (Oct., 2017).
\newblock \doi{10.1103/PhysRevLett.119.161101}.

\bibitem{rui+18}
M.~{Ruiz}, S.~L. {Shapiro}, and A.~{Tsokaros}, {GW170817, General Relativistic
  Magnetohydrodynamic Simulations, and the Neutron Star Maximum Mass},
  \emph{Phys. Rev. D}. {\bf 97}, \penalty0 021501  (Jan., 2018).

\bibitem{mas+16}
A.~{Maselli}, H.~O. {Silva}, M.~{Minamitsuji}, and E.~{Berti}, {Neutron stars
  in Horndeski gravity}, \emph{Phys. Rev. D}. 93\penalty0 (12):\penalty0 124056
   (June, 2016).
\newblock \doi{10.1103/PhysRevD.93.124056}.

\bibitem{pan+11}
P.~{Pani}, E.~{Berti}, V.~{Cardoso}, and J.~{Read}, {Compact stars in
  alternative theories of gravity: Einstein-Dilaton-Gauss-Bonnet gravity},
  \emph{Phys. Rev. D}. 84\penalty0 (10):\penalty0 104035  (Nov., 2011).
\newblock \doi{10.1103/PhysRevD.84.104035}.

\bibitem{kle+16}
B.~{Kleihaus}, J.~{Kunz}, S.~{Mojica}, and M.~{Zagermann}, {Rapidly rotating
  neutron stars in dilatonic Einstein-Gauss-Bonnet theory}, \emph{Phys. Rev.
  D}. 93\penalty0 (6):\penalty0 064077  (Mar., 2016).
\newblock \doi{10.1103/PhysRevD.93.064077}.

\bibitem{yag+13}
K.~{Yagi}, L.~C. {Stein}, N.~{Yunes}, and T.~{Tanaka}, {Isolated and binary
  neutron stars in dynamical Chern-Simons gravity}, \emph{Phys. Rev. D}.
  87\penalty0 (8):\penalty0 084058  (Apr., 2013).
\newblock \doi{10.1103/PhysRevD.87.084058}.

\bibitem{coo+10}
A.~{Cooney}, S.~{Dedeo}, and D.~{Psaltis}, {Neutron stars in $f(R)$ gravity
  with perturbative constraints}, \emph{Phys. Rev. D}. 82\penalty0
  (6):\penalty0 064033  (Sept., 2010).
\newblock \doi{10.1103/PhysRevD.82.064033}.

\bibitem{ara+11}
S.~{Arapo\u{g}lu}, C.~{Deliduman}, and K.~Y. {Ek{\c{s}}i}, {Constraints on
  perturbative $f(R)$ gravity via neutron stars}, \emph{J. Cosm. Astropart.
  Phys.} 7:\penalty0 020  (July, 2011).
\newblock \doi{10.1088/1475-7516/2011/07/020}.

\bibitem{del+12}
C.~{Deliduman}, K.~Y. {Ek{\c s}i}, and V.~{Kele{\c s}}, {Neutron star solutions
  in perturbative quadratic gravity}, \emph{J. Cosm. Astropart. Phys.}
  5:\penalty0 036  (May, 2012).
\newblock \doi{10.1088/1475-7516/2012/05/036}.

\bibitem{ant+13}
J.~{Antoniadis}, P.~C.~C. {Freire}, N.~{Wex}, T.~M. {Tauris}, R.~S. {Lynch},
  M.~H. {van Kerkwijk}, M.~{Kramer}, C.~{Bassa}, V.~S. {Dhillon}, T.~{Driebe},
  J.~W.~T. {Hessels}, V.~M. {Kaspi}, V.~I. {Kondratiev}, N.~{Langer}, T.~R.
  {Marsh}, M.~A. {McLaughlin}, T.~T. {Pennucci}, S.~M. {Ransom}, I.~H.
  {Stairs}, J.~{van Leeuwen}, J.~P.~W. {Verbiest}, and D.~G. {Whelan}, {A
  Massive Pulsar in a Compact Relativistic Binary}, \emph{Science}. {\bf 340},
  \penalty0 448  (Apr., 2013).
\newblock \doi{10.1126/science.1233232}.

\bibitem{dem+10}
P.~B. {Demorest}, T.~{Pennucci}, S.~M. {Ransom}, M.~S.~E. {Roberts}, and
  J.~W.~T. {Hessels}, {A two-solar-mass neutron star measured using Shapiro
  delay}, \emph{Nature}. {\bf 467}, \penalty0 1081--1083  (Oct., 2010).
\newblock \doi{10.1038/nature09466}.

\bibitem{ore+13}
M.~{Orellana}, F.~{Garc{\'{\i}}a}, F.~A. {Teppa Pannia}, and G.~E. {Romero},
  {Structure of neutron stars in R-squared gravity}, \emph{Gen. Relativ.
  Gravit.} {\bf 45}, \penalty0 771--783  (Apr., 2013).
\newblock \doi{10.1007/s10714-013-1501-5}.

\bibitem{yaz+14}
S.~S. {Yazadjiev}, D.~D. {Doneva}, K.~D. {Kokkotas}, and K.~V. {Staykov},
  {Non-perturbative and self-consistent models of neutron stars in R-squared
  gravity}, \emph{J. Cosm. Astropart. Phys.} 6:\penalty0 003  (June, 2014).
\newblock \doi{10.1088/1475-7516/2014/06/003}.

\bibitem{bla+18}
J.~L. {Bl{\'a}zquez-Salcedo}, D.~D. {Doneva}, J.~{Kunz}, K.~V. {Staykov}, and
  S.~S. {Yazadjiev}, {Axial quasi-normal modes of neutron stars in $R^2$
  gravity}, \emph{ArXiv e-prints}  (Apr., 2018).

\bibitem{ast+15}
A.~V. {Astashenok}, S.~{Capozziello}, and S.~D. {Odintsov}, {Nonperturbative
  models of quark stars in $f(R)$ gravity}, \emph{Phys. Lett. B}. {\bf 742},
  \penalty0 160--166  (Mar., 2015).
\newblock \doi{10.1016/j.physletb.2015.01.030}.

\bibitem{cap+16}
S.~{Capozziello}, M.~{De Laurentis}, R.~{Farinelli}, and S.~D. {Odintsov},
  {Mass-radius relation for neutron stars in $f(R)$ gravity}, \emph{Phys. Rev.
  D}. 93\penalty0 (2):\penalty0 023501  (Jan., 2016).
\newblock \doi{10.1103/PhysRevD.93.023501}.

\bibitem{apa+16}
M.~{Aparicio Resco}, {\'A}.~{de la Cruz-Dombriz}, F.~J. {Llanes Estrada}, and
  V.~{Zapatero Castrillo}, {On neutron stars in $f(R)$ theories: Small radii,
  large masses and large energy emitted in a merger}, \emph{Physics of the Dark
  Universe}. {\bf 13}, \penalty0 147--161  (Sept., 2016).
\newblock \doi{10.1016/j.dark.2016.07.001}.

\bibitem{kai+07}
K.~{Kainulainen}, V.~{Reijonen}, and D.~{Sunhede}, {Interior spacetimes of
  stars in Palatini $f(R)$ gravity}, \emph{Phys. Rev. D}. 76\penalty0
  (4):\penalty0 043503  (Aug., 2007).
\newblock \doi{10.1103/PhysRevD.76.043503}.

\bibitem{liu+18}
T.~{Liu}, X.~{Zhang}, and W.~{Zhao}, {Constraining $f(R)$ gravity in solar
  system, cosmology and binary pulsar systems}, \emph{Phys. Lett. B}. {\bf
  777}, \penalty0 286--293  (Jan., 2018).

\bibitem{bar+08}
E.~{Barausse}, T.~P. {Sotiriou}, and J.~C. {Miller}, {Curvature singularities,
  tidal forces and the viability of Palatini $f(R)$ gravity}, \emph{Class.
  Quantum Grav.} 25\penalty0 (10):\penalty0 105008  (May, 2008).
\newblock \doi{10.1088/0264-9381/25/10/105008}.

\bibitem{barf+08}
E.~{Barausse}, T.~P. {Sotiriou}, and J.~C. {Miller}, {FAST TRACK COMMUNICATION:
  A no-go theorem for polytropic spheres in Palatini $f(R)$ gravity},
  \emph{Class. Quantum Grav.} 25\penalty0 (6):\penalty0 062001  (Mar., 2008).
\newblock \doi{10.1088/0264-9381/25/6/062001}.

\bibitem{tep+17}
F.~A. {Teppa Pannia}, F.~{Garc{\'{\i}}a}, S.~E. {Perez Bergliaffa},
  M.~{Orellana}, and G.~E. {Romero}, {Structure of compact stars in R-squared
  Palatini gravity}, \emph{Gen. Relativ. Gravit.} 49:\penalty0 25  (Feb.,
  2017).
\newblock \doi{10.1007/s10714-016-2182-7}.

\bibitem{fol18}
V.~{Folomeev}, {Anisotropic neutron stars in R$^{2}$ gravity}, \emph{Phys. Rev.
  D}. 97\penalty0 (12):\penalty0 124009  (June, 2018).
\newblock \doi{10.1103/PhysRevD.97.124009}.

\bibitem{sta+14}
K.~V. {Staykov}, D.~D. {Doneva}, S.~S. {Yazadjiev}, and K.~D. {Kokkotas},
  {Slowly rotating neutron and strange stars in R$^{2}$ gravity}, \emph{J.
  Cosm. Astropart. Phys.} 10:\penalty0 006  (Oct., 2014).
\newblock \doi{10.1088/1475-7516/2014/10/006}.

\bibitem{yaz+15}
S.~S. {Yazadjiev}, D.~D. {Doneva}, and K.~D. {Kokkotas}, {Rapidly rotating
  neutron stars in R -squared gravity}, \emph{Phys. Rev. D}. 91\penalty0
  (8):\penalty0 084018  (Apr., 2015).
\newblock \doi{10.1103/PhysRevD.91.084018}.

\bibitem{che+13}
M.-K. {Cheoun}, C.~{Deliduman}, C.~{G{\"u}ng{\"o}r}, V.~{Kele{\c s}}, C.~Y.
  {Ryu}, T.~{Kajino}, and G.~J. {Mathews}, {Neutron stars in a perturbative
  $f(R)$ gravity model with strong magnetic fields}, \emph{J. Cosm. Astropart.
  Phys.} 10:\penalty0 021  (Oct., 2013).
\newblock \doi{10.1088/1475-7516/2013/10/021}.

\bibitem{ast+15b}
A.~V. {Astashenok}, S.~{Capozziello}, and S.~D. {Odintsov}, {Extreme neutron
  stars from Extended Theories of Gravity}, \emph{J. Cosm. Astropart. Phys.}
  1:\penalty0 001  (Jan., 2015).
\newblock \doi{10.1088/1475-7516/2015/01/001}.

\bibitem{bak+16}
E.~{Bakirova} and V.~{Folomeev}, {Dipole magnetic field of neutron stars in
  $f(R)$ gravity}, \emph{Gen. Relativ. Gravit.} 48:\penalty0 135  (Oct., 2016).
\newblock \doi{10.1007/s10714-016-2127-1}.

\bibitem{ala+13}
H.~{Alavirad} and J.~M. {Weller}, {Modified gravity with logarithmic curvature
  corrections and the structure of relativistic stars}, \emph{Phys. Rev. D}.
  88\penalty0 (12):\penalty0 124034  (Dec., 2013).
\newblock \doi{10.1103/PhysRevD.88.124034}.

\bibitem{lop+17}
F.~G. {Lopez Armengol} and G.~E. {Romero}, {Neutron stars in
  Scalar-Tensor-Vector Gravity}, \emph{Gen. Relativ. Gravit.} 49:\penalty0 27
  (Feb., 2017).
\newblock \doi{10.1007/s10714-017-2184-0}.

\bibitem{sil+04}
R.~R. {Silbar} and S.~{Reddy}, {Neutron stars for undergraduates}, \emph{Am. J.
  Phys.} {\bf 72}, \penalty0 892--905  (July, 2004).
\newblock \doi{10.1119/1.1703544}.

\bibitem{dou+01}
F.~{Douchin} and P.~{Haensel}, {A unified equation of state of dense matter and
  neutron star structure}, \emph{Astron. Astroph.} {\bf 380}, \penalty0
  151--167  (Dec., 2001).
\newblock \doi{10.1051/0004-6361:20011402}.

\bibitem{pan+89}
V.~R. {Pandharipande} and D.~G. {Ravenhall}.
\newblock {Hot Nuclear Matter}.
\newblock In eds. M.~{Soyeur}, H.~{Flocard}, B.~{Tamain}, and M.~{Porneuf},
  \emph{NATO Advanced Science Institutes (ASI) Series B}, vol. 205, \emph{NATO
  Advanced Science Institutes (ASI) Series B}, p. 103  (1989).

\bibitem{gor+10}
S.~{Goriely}, N.~{Chamel}, and J.~M. {Pearson}, {Further explorations of
  Skyrme-Hartree-Fock-Bogoliubov mass formulas. XII. Stiffness and stability of
  neutron-star matter}, \emph{Phys. Rev. C}. 82\penalty0 (3):\penalty0 035804
  (Sept., 2010).
\newblock \doi{10.1103/PhysRevC.82.035804}.

\bibitem{pea+11}
J.~M. {Pearson}, S.~{Goriely}, and N.~{Chamel}, {Properties of the outer crust
  of neutron stars from Hartree-Fock-Bogoliubov mass models}, \emph{Phys. Rev.
  C}. 83\penalty0 (6):\penalty0 065810  (June, 2011).
\newblock \doi{10.1103/PhysRevC.83.065810}.

\bibitem{pea+12}
J.~M. {Pearson}, N.~{Chamel}, S.~{Goriely}, and C.~{Ducoin}, {Inner crust of
  neutron stars with mass-fitted Skyrme functionals}, \emph{Phys. Rev. C}.
  85\penalty0 (6):\penalty0 065803  (June, 2012).
\newblock \doi{10.1103/PhysRevC.85.065803}.

\bibitem{kiz+13}
B.~{Kiziltan}, A.~{Kottas}, M.~{De Yoreo}, and S.~E. {Thorsett}, {The Neutron
  Star Mass Distribution}, \emph{Astrophys. J.} 778:\penalty0 66  (Nov., 2013).
\newblock \doi{10.1088/0004-637X/778/1/66}.

\bibitem{oze+12}
F.~{\"{O}zel}, D.~{Psaltis}, R.~{Narayan}, and A.~{Santos Villarreal}, {On the
  Mass Distribution and Birth Masses of Neutron Stars}, \emph{Astrophys. J.}
  757:\penalty0 55  (Sept., 2012).
\newblock \doi{10.1088/0004-637X/757/1/55}.

\bibitem{sch16}
K.~{Schwarzschild}, {{\"U}ber das Gravitationsfeld eines Massenpunktes nach der
  Einsteinschen Theorie}, \emph{Sitzungsberichte der K{\"o}niglich
  Preu{\ss}ischen Akademie der Wissenschaften (Berlin), 1916, Seite 189-196}
  (1916).

\bibitem{haz+63}
C.~{Hazard}, M.~B. {Mackey}, and A.~J. {Shimmins}, {Investigation of the Radio
  Source 3C 273 By The Method of Lunar Occultations}, \emph{Nature}. {\bf 197},
  \penalty0 1037--1039  (Mar., 1963).
\newblock \doi{10.1038/1971037a0}.

\bibitem{web+72}
B.~L. {Webster} and P.~{Murdin}, {Cygnus X-1-a Spectroscopic Binary with a
  Heavy Companion?}, \emph{Nature}. {\bf 235}, \penalty0 37--38  (Jan., 1972).
\newblock \doi{10.1038/235037a0}.

\bibitem{bol72}
C.~T. {Bolton}, {Dimensions of the Binary System HDE 226868 = Cygnus X-1},
  \emph{Nature}. {\bf 240}, \penalty0 124--127  (Dec., 1972).
\newblock \doi{10.1038/physci240124a0}.

\bibitem{abb+16c}
B.~P. {Abbott}, R.~{Abbott}, T.~D. {Abbott}, M.~R. {Abernathy}, F.~{Acernese},
  K.~{Ackley}, C.~{Adams}, T.~{Adams}, P.~{Addesso}, R.~X. {Adhikari}, and
  et~al., {GW151226: Observation of Gravitational Waves from a 22-Solar-Mass
  Binary Black Hole Coalescence}, \emph{Phys. Rev. Lett.} 116\penalty0
  (24):\penalty0 241103  (June, 2016).
\newblock \doi{10.1103/PhysRevLett.116.241103}.

\bibitem{abb+17a}
B.~P. {Abbott}, R.~{Abbott}, T.~D. {Abbott}, F.~{Acernese}, K.~{Ackley},
  C.~{Adams}, T.~{Adams}, P.~{Addesso}, R.~X. {Adhikari}, V.~B. {Adya}, and
  et~al., {GW170104: Observation of a 50-Solar-Mass Binary Black Hole
  Coalescence at Redshift 0.2}, \emph{Phys. Rev. Lett.} 118\penalty0
  (22):\penalty0 221101  (June, 2017).
\newblock \doi{10.1103/PhysRevLett.118.221101}.

\bibitem{abb+17e}
B.~P. {Abbott}, R.~{Abbott}, T.~D. {Abbott}, F.~{Acernese}, K.~{Ackley},
  C.~{Adams}, T.~{Adams}, P.~{Addesso}, R.~X. {Adhikari}, V.~B. {Adya}, and
  et~al., {GW170608: Observation of a 19 Solar-mass Binary Black Hole
  Coalescence}, \emph{Astrophys. J. Lett.} 851:\penalty0 L35  (Dec., 2017).
\newblock \doi{10.3847/2041-8213/aa9f0c}.

\bibitem{abb+17a1}
B.~P. {Abbott}, R.~{Abbott}, T.~D. {Abbott}, F.~{Acernese}, K.~{Ackley},
  C.~{Adams}, T.~{Adams}, P.~{Addesso}, R.~X. {Adhikari}, V.~B. {Adya}, and
  et~al., {GW170814: A Three-Detector Observation of Gravitational Waves from a
  Binary Black Hole Coalescence}, \emph{Phys. Rev. Lett.} 119\penalty0
  (14):\penalty0 141101  (Oct., 2017).
\newblock \doi{10.1103/PhysRevLett.119.141101}.

\bibitem{isr67}
W.~{Israel}, {Event Horizons in Static Vacuum Space-Times}, \emph{Phys. Rev.}
  {\bf 164}, \penalty0 1776--1779  (Dec., 1967).
\newblock \doi{10.1103/PhysRev.164.1776}.

\bibitem{isr68}
W.~{Israel}, {Event horizons in static electrovac space-times}, \emph{Commun.
  Math. Phys.} {\bf 8}, \penalty0 245--260  (Sept., 1968).
\newblock \doi{10.1007/BF01645859}.

\bibitem{car71}
B.~{Carter}, {Axisymmetric Black Hole Has Only Two Degrees of Freedom},
  \emph{Phys. Rev. Lett.} {\bf 26}, \penalty0 331--333  (Feb., 1971).
\newblock \doi{10.1103/PhysRevLett.26.331}.

\bibitem{hawk72}
S.~W. {Hawking}, {Black holes in general relativity}, \emph{Commun. Math.
  Phys.} {\bf 25}, \penalty0 152--166  (June, 1972).
\newblock \doi{10.1007/BF01877517}.

\bibitem{rob75}
D.~C. {Robinson}, {Uniqueness of the Kerr black hole}, \emph{Phys. Rev. Lett.}
  {\bf 34}, \penalty0 905  (Apr., 1975).
\newblock \doi{10.1103/PhysRevLett.34.905}.

\bibitem{sot+10}
T.~P. {Sotiriou} and V.~{Faraoni}, {$f(R)$ theories of gravity}, \emph{Rev.
  Mod. Phys.} {\bf 82}, \penalty0 451--497  (Jan., 2010).
\newblock \doi{10.1103/RevModPhys.82.451}.

\bibitem{far10}
V.~{Faraoni}, {Jebsen-Birkhoff theorem in alternative gravity}, \emph{Phys.
  Rev. D}. 81\penalty0 (4):\penalty0 044002  (Feb., 2010).
\newblock \doi{10.1103/PhysRevD.81.044002}.

\bibitem{can+16}
P.~{Ca\~{n}ate}, L.~G. {Jaime}, and M.~{Salgado}, {Spherically symmetric black
  holes in $f(R)$ gravity: is geometric scalar hair supported?}, \emph{Class.
  Quantum Grav.} 33\penalty0 (15):\penalty0 155005  (Aug., 2016).
\newblock \doi{10.1088/0264-9381/33/15/155005}.

\bibitem{can18}
P.~{Ca\~{n}ate}, {A no-hair theorem for black holes in $f(R)$ gravity},
  \emph{Class. Quantum Grav.} 35\penalty0 (2):\penalty0 025018  (Jan., 2018).
\newblock \doi{10.1088/1361-6382/aa8e2e}.

\bibitem{olm+11}
G.~J. {Olmo} and D.~{Rubiera-Garcia}, {Palatini $f(R)$ black holes in nonlinear
  electrodynamics}, \emph{Phys. Rev. D}. 84\penalty0 (12):\penalty0 124059
  (Dec., 2011).
\newblock \doi{10.1103/PhysRevD.84.124059}.

\bibitem{olm+12a}
G.~J. {Olmo} and D.~{Rubiera-Garcia}, {Reissner-Nordstr\"{o}m black holes in
  extended Palatini theories}, \emph{Phys. Rev. D}. 86\penalty0 (4):\penalty0
  044014  (Aug., 2012).
\newblock \doi{10.1103/PhysRevD.86.044014}.

\bibitem{olm+12b}
G.~J. {Olmo} and D.~{Rubiera-Garcia}, {Nonsingular black holes in quadratic
  Palatini gravity}, \emph{Eur. Phys. J. C}. 72:\penalty0 2098  (Aug., 2012).
\newblock \doi{10.1140/epjc/s10052-012-2098-7}.

\bibitem{baz+14}
D.~{Bazeia}, L.~{Losano}, G.~J. {Olmo}, and D.~{Rubiera-Garcia}, {Black holes
  in five-dimensional Palatini $f(R)$ gravity and implications for the AdS/CFT
  correspondence}, \emph{Phys. Rev. D}. 90\penalty0 (4):\penalty0 044011
  (Aug., 2014).
\newblock \doi{10.1103/PhysRevD.90.044011}.

\bibitem{Olm+15a}
G.~J. {Olmo}, D.~{Rubiera-Garcia}, and A.~{Sanchez-Puente}.
\newblock {Geometric aspects of charged black holes in Palatini theories}.
\newblock In \emph{Journal of Physics Conference Series}, vol. 600,
  \emph{Journal of Physics Conference Series}, p. 012042  (Apr., 2015).
\newblock \doi{10.1088/1742-6596/600/1/012042}.

\bibitem{psa+08}
D.~{Psaltis}, D.~{Perrodin}, K.~R. {Dienes}, and I.~{Mocioiu}, {Kerr Black
  Holes Are Not Unique to General Relativity}, \emph{Phys. Rev. Lett.}
  100\penalty0 (9):\penalty0 091101  (Mar., 2008).
\newblock \doi{10.1103/PhysRevLett.100.091101}.

\bibitem{bar+08a}
E.~{Barausse} and T.~P. {Sotiriou}, {Perturbed Kerr Black Holes Can Probe
  Deviations from General Relativity}, \emph{Phys. Rev. Lett.} 101\penalty0
  (9):\penalty0 099001  (Aug., 2008).
\newblock \doi{10.1103/PhysRevLett.101.099001}.

\bibitem{del+09}
A.~{de La Cruz-Dombriz}, A.~{Dobado}, and A.~L. {Maroto}, {Black holes in
  $f(R)$ theories}, \emph{Phys. Rev. D}. 80\penalty0 (12):\penalty0 124011
  (Dec., 2009).
\newblock \doi{10.1103/PhysRevD.80.124011}.

\bibitem{del+11}
A.~{de La Cruz-Dombriz}, A.~{Dobado}, and A.~L. {Maroto}, {Erratum: Black holes
  in $f(R)$ theories [Phys. Rev. DPRVDAQ1550-7998 80, 124011 (2009)]},
  \emph{Phys. Rev. D}. 83\penalty0 (2):\penalty0 029903  (Jan., 2011).
\newblock \doi{10.1103/PhysRevD.83.029903}.

\bibitem{moo+11}
T.~{Moon}, Y.~S. {Myung}, and E.~J. {Son}, {$f(R)$ black holes}, \emph{Gen.
  Relativ. Gravit.} {\bf 43}, \penalty0 3079--3098  (Nov., 2011).
\newblock \doi{10.1007/s10714-011-1225-3}.

\bibitem{hen+12}
S.~H. {Hendi}, B.~E. {Panah}, and S.~M. {Mousavi}, {Some exact solutions of
  $f(R)$ gravity with charged (a)dS black hole interpretation}, \emph{Gen.
  Relativ. Gravit.} {\bf 44}, \penalty0 835--853  (Apr., 2012).
\newblock \doi{10.1007/s10714-011-1307-2}.

\bibitem{cem+11}
J.~A.~R. {Cembranos}, A.~{de la Cruz-Dombriz}, and P.~{Jimeno Romero},
  {Kerr-Newman black holes in $f(R)$ theories}, \emph{ArXiv e-prints}  (Sept.,
  2011).

\bibitem{maz+12}
S.~{Habib Mazharimousavi}, M.~{Halilsoy}, and T.~{Tahamtan}, {Solutions for
  $f(R)$ gravity coupled with electromagnetic field}, \emph{Eur. Phys. J. C}.
  72:\penalty0 1851  (Jan., 2012).
\newblock \doi{10.1140/epjc/s10052-011-1851-7}.

\bibitem{maz+12b}
S.~{Habib Mazharimousavi}, M.~{Halilsoy}, and T.~{Tahamtan}, {Constant
  curvature $f(R)$ gravity minimally coupled with Yang-Mills field}, \emph{Eur.
  Phys. J. C}. 72:\penalty0 1958  (Mar., 2012).
\newblock \doi{10.1140/epjc/s10052-012-1958-5}.

\bibitem{she12}
A.~{Sheykhi}, {Higher-dimensional charged $f(R)$ black holes}, \emph{Phys. Rev.
  D}. 86\penalty0 (2):\penalty0 024013  (July, 2012).
\newblock \doi{10.1103/PhysRevD.86.024013}.

\bibitem{ber+11}
S.~E.~P. {Bergliaffa} and Y.~E.~C.~D.~O. {Nunes}, {Static and spherically
  symmetric black holes in $f(R)$ theories}, \emph{Phys. Rev. D}. 84\penalty0
  (8):\penalty0 084006  (Oct., 2011).
\newblock \doi{10.1103/PhysRevD.84.084006}.

\bibitem{maz+13}
S.~H. {Mazharimousavi}, M.~{Kerachian}, and M.~{Halilsoy}, {Existence of
  Reissner-Nordstr{\"o}m-Type Black Holes in $f(R)$ Gravity}, \emph{Int. J.
  Mod. Phys. D}. 22:\penalty0 1350057  (July, 2013).
\newblock \doi{10.1142/S0218271813500570}.

\bibitem{ami+16}
Z.~{Amirabi}, M.~{Halilsoy}, and S.~H. {Mazharimousavi}, {Generation of
  spherically symmetric metrics in $f(R)$ gravity}, \emph{Eur. Phys. J. C}.
  76:\penalty0 338  (June, 2016).
\newblock \doi{10.1140/epjc/s10052-016-4164-z}.

\bibitem{gao+16}
C.~{Gao} and Y.-G. {Shen}, {Exact solutions in $f(R)$ theory of gravity},
  \emph{Gen. Relativ. Gravit.} 48:\penalty0 131  (Oct., 2016).
\newblock \doi{10.1007/s10714-016-2128-0}.

\bibitem{hen+11}
S.~H. {Hendi} and D.~{Momeni}, {Black-hole solutions in $f(R)$ gravity with
  conformal anomaly}, \emph{Eur. Phys. J. C}. 71:\penalty0 1823  (Dec., 2011).
\newblock \doi{10.1140/epjc/s10052-011-1823-y}.

\bibitem{rod+16}
M.~E. {Rodrigues}, E.~L.~B. {Junior}, G.~T. {Marques}, and V.~T. {Zanchin},
  {Regular black holes in $f(R)$ gravity coupled to nonlinear electrodynamics},
  \emph{Phys. Rev. D}. 94\penalty0 (2):\penalty0 024062  (July, 2016).
\newblock \doi{10.1103/PhysRevD.94.024062}.

\bibitem{myu+11}
Y.~S. {Myung}, T.~{Moon}, and E.~J. {Son}, {Stability of $f(R)$ black holes},
  \emph{Phys. Rev. D}. 83\penalty0 (12):\penalty0 124009  (June, 2011).
\newblock \doi{10.1103/PhysRevD.83.124009}.

\bibitem{moo+11b}
T.~{Moon}, Y.~S. {Myung}, and E.~J. {Son}, {Stability analysis of $f(R)$-AdS
  black holes}, \emph{Eur. Phys. J. C}. 71:\penalty0 1777  (Oct., 2011).
\newblock \doi{10.1140/epjc/s10052-011-1777-0}.

\bibitem{myu11}
Y.~S. {Myung}, {Instability of a rotating black hole in a limited form of
  $f(R)$ gravity}, \emph{Phys. Rev. D}. 84\penalty0 (2):\penalty0 024048
  (July, 2011).
\newblock \doi{10.1103/PhysRevD.84.024048}.

\bibitem{car73}
B.~{Carter}.
\newblock {Black hole equilibrium states.}
\newblock In eds. C.~{Dewitt} and B.~S. {Dewitt}, \emph{Black Holes (Les Astres
  Occlus)}, pp. 57--214  (1973).

\bibitem{cal+18}
M.~{Calz{\`a}}, M.~{Rinaldi}, and L.~{Sebastiani}, {A special class of
  solutions in F( R)-gravity}, \emph{European Physical Journal C}. 78\penalty0
  (3):\penalty0 178  (Mar., 2018).
\newblock \doi{10.1140/epjc/s10052-018-5681-8}.

\bibitem{gou+11}
L.~{Gou}, J.~E. {McClintock}, M.~J. {Reid}, J.~A. {Orosz}, J.~F. {Steiner},
  R.~{Narayan}, J.~{Xiang}, R.~A. {Remillard}, K.~A. {Arnaud}, and S.~W.
  {Davis}, {The Extreme Spin of the Black Hole in Cygnus X-1}, \emph{Astrophys.
  J.} 742:\penalty0 85  (Dec., 2011).
\newblock \doi{10.1088/0004-637X/742/2/85}.

\bibitem{per+13}
D.~{P\'{e}rez}, G.~E. {Romero}, and S.~E. {Perez Bergliaffa}, {Accretion disks
  around black holes in modified strong gravity}, \emph{Astron. Astroph.}
  551:\penalty0 A4  (Mar., 2013).
\newblock \doi{10.1051/0004-6361/201220378}.

\bibitem{mof+15a}
J.~W. {Moffat}, {Black holes in modified gravity (MOG)}, \emph{Eur. Phys. J.
  C}. 75:\penalty0 175  (Apr., 2015).
\newblock \doi{10.1140/epjc/s10052-015-3405-x}.

\bibitem{mof+13}
J.~W. {Moffat} and S.~{Rahvar}, {The MOG weak field approximation and
  observational test of galaxy rotation curves}, \emph{Mon. Not. R. Astron.
  Soc.} {\bf 436}, \penalty0 1439--1451  (Dec., 2013).
\newblock \doi{10.1093/mnras/stt1670}.

\bibitem{mof+14}
J.~W. {Moffat} and S.~{Rahvar}, {The MOG weak field approximation - II.
  Observational test of Chandra X-ray clusters}, \emph{Mon. Not. R. Astron.
  Soc.} {\bf 441}, \penalty0 3724--3732  (July, 2014).
\newblock \doi{10.1093/mnras/stu855}.

\bibitem{mof+15}
J.~W. {Moffat} and V.~T. {Toth}, {Rotational velocity curves in the Milky Way
  as a test of modified gravity}, \emph{Phys. Rev. D}. 91\penalty0
  (4):\penalty0 043004  (Feb., 2015).
\newblock \doi{10.1103/PhysRevD.91.043004}.

\bibitem{sha72}
N.~I. {Shakura}, {Disk Model of Gas Accretion on a Relativistic Star in a Close
  Binary System.}, \emph{Astronomicheskii Zhurnal}. {\bf 49}, \penalty0 921
  (Oct., 1972).

\bibitem{sha+73}
N.~I. Shakura and R.~A. Sunyaev, Black holes in binary systems. observational
  appearance, \emph{Astron. Astroph.} {\bf 24}, \penalty0 337--355  (1973).

\bibitem{nov+73}
I.~D. {Novikov} and K.~S. {Thorne}.
\newblock {Astrophysics of black holes.}
\newblock In ed. {C.~Dewitt \& B.~S.~Dewitt}, \emph{Black Holes (Les Astres
  Occlus)}, pp. 343--450  (1973).

\bibitem{pag+74}
D.~N. {Page} and K.~S. {Thorne}, {Disk-Accretion onto a Black Hole.
  Time-Averaged Structure of Accretion Disk}, \emph{Astrophys. J.} {\bf 191},
  \penalty0 499--506  (July, 1974).
\newblock \doi{10.1086/152990}.

\bibitem{har+09}
T.~{Harko}, Z.~{Kov{\'a}cs}, and F.~S.~N. {Lobo}, {Thin accretion disks in
  stationary axisymmetric wormhole spacetimes}, \emph{Phys. Rev. D}.
  79\penalty0 (6):\penalty0 064001  (Mar., 2009).
\newblock \doi{10.1103/PhysRevD.79.064001}.

\bibitem{stu+99}
Z.~{Stuchl{\'{\i}}k} and S.~{Hled{\'{\i}}k}, {Some properties of the
  Schwarzschild-de Sitter and Schwarzschild-anti-de Sitter spacetimes},
  \emph{Phys. Rev. D}. 60\penalty0 (4):\penalty0 044006  (Aug., 1999).
\newblock \doi{10.1103/PhysRevD.60.044006}.

\bibitem{rez+03}
L.~{Rezzolla}, O.~{Zanotti}, and J.~A. {Font}, {Dynamics of thick discs around
  Schwarzschild-de Sitter black holes}, \emph{Astron. Astroph.} {\bf 412},
  \penalty0 603--613  (Dec., 2003).
\newblock \doi{10.1051/0004-6361:20031457}.

\bibitem{stu+04}
Z.~{Stuchl{\'{\i}}k} and P.~{Slan{\'y}}, {Equatorial circular orbits in the
  Kerr de Sitter spacetimes}, \emph{Phys. Rev. D}. 69\penalty0 (6):\penalty0
  064001  (Mar., 2004).
\newblock \doi{10.1103/PhysRevD.69.064001}.

\bibitem{sla+13}
P.~{Slan{\'y}}, M.~{Pokorn{\'a}}, and Z.~{Stuchl{\'{\i}}k}, {Equatorial
  circular orbits in Kerr-anti-de Sitter spacetimes}, \emph{Gen. Relativ.
  Gravit.} {\bf 45}, \penalty0 2611--2633  (Dec., 2013).
\newblock \doi{10.1007/s10714-013-1606-x}.

\bibitem{per+17}
D.~{P{\'e}rez}, F.~G.~L. {Armengol}, and G.~E. {Romero}, {Accretion disks
  around black holes in scalar-tensor-vector gravity}, \emph{Phys. Rev. D}.
  95\penalty0 (10):\penalty0 104047  (May, 2017).
\newblock \doi{10.1103/PhysRevD.95.104047}.

\bibitem{mof+08}
J.~W. {Moffat} and V.~T. {Toth}, {Testing Modified Gravity with Globular
  Cluster Velocity Dispersions}, \emph{Astrophys. J.} 680:\penalty0 1158-1161
  (June, 2008).
\newblock \doi{10.1086/587926}.

\bibitem{bro+06}
J.~R. {Brownstein} and J.~W. {Moffat}, {Galaxy cluster masses without
  non-baryonic dark matter}, \emph{Mon. Not. R. Astron. Soc.} {\bf 367},
  \penalty0 527--540  (Apr., 2006).
\newblock \doi{10.1111/j.1365-2966.2006.09996.x}.

\bibitem{oro+11}
J.~A. {Orosz}, J.~E. {McClintock}, J.~P. {Aufdenberg}, R.~A. {Remillard}, M.~J.
  {Reid}, R.~{Narayan}, and L.~{Gou}, {The Mass of the Black Hole in Cygnus
  X-1}, \emph{Astrophys. J.} 742:\penalty0 84  (Dec., 2011).
\newblock \doi{10.1088/0004-637X/742/2/84}.

\bibitem{ahm+16}
A.~K. {Ahmed}, M.~{Azreg-A{\"i}nou}, M.~{Faizal}, and M.~{Jamil}, {Cyclic and
  heteroclinic flows near general static spherically symmetric black holes},
  \emph{Eur. Phys. J. C}. 76:\penalty0 280  (May, 2016).
\newblock \doi{10.1140/epjc/s10052-016-4112-y}.

\bibitem{azr17}
M.~{Azreg-A{\"i}nou}, {Cyclic and heteroclinic flows near general static
  spherically symmetric black holes: semi-cyclic flows - addendum and
  corrigendum}, \emph{Eur. Phys. J. C}. 77:\penalty0 36  (Jan., 2017).
\newblock \doi{10.1140/epjc/s10052-017-4613-3}.

\bibitem{ali+16}
N.~{Alipour}, A.~R. {Khesali}, and K.~{Nozari}, {Dynamics of accretion disks in
  a constant curvature $f(R)$-gravity}, \emph{Astrophys. Space Sci.}
  361:\penalty0 240  (July, 2016).
\newblock \doi{10.1007/s10509-016-2829-6}.

\bibitem{bha+15}
C.~{Bhattacharjee}, R.~{Das}, and S.~M. {Mahajan}, {Novel mechanism for
  vorticity generation in black-hole accretion disks}, \emph{Phys. Rev. D}.
  91\penalty0 (12):\penalty0 123005  (June, 2015).
\newblock \doi{10.1103/PhysRevD.91.123005}.

\bibitem{lop+17a}
F.~G. {Lopez Armengol} and G.~E. {Romero}, {Effects of Scalar-Tensor-Vector
  Gravity on relativistic jets}, \emph{Astrophys. Space Sci.} 362:\penalty0 214
   (Nov., 2017).
\newblock \doi{10.1007/s10509-017-3197-6}.

\bibitem{geb+11}
K.~{Gebhardt}, J.~{Adams}, D.~{Richstone}, T.~R. {Lauer}, S.~M. {Faber},
  K.~{G{\"u}ltekin}, J.~{Murphy}, and S.~{Tremaine}, {The Black Hole Mass in
  M87 from Gemini/NIFS Adaptive Optics Observations}, \emph{Astrophys. J.}
  729:\penalty0 119  (Mar., 2011).
\newblock \doi{10.1088/0004-637X/729/2/119}.

\bibitem{li+09}
Y.-R. {Li}, Y.-F. {Yuan}, J.-M. {Wang}, J.-C. {Wang}, and S.~{Zhang}, {Spins of
  Supermassive Black Holes in M87. II. Fully General Relativistic
  Calculations}, \emph{Astrophys. J.} {\bf 699}, \penalty0 513--524  (July,
  2009).
\newblock \doi{10.1088/0004-637X/699/1/513}.

\bibitem{mer+16}
F.~{Mertens}, A.~P. {Lobanov}, R.~C. {Walker}, and P.~E. {Hardee}, {Kinematics
  of the jet in M 87 on scales of 100-1000 Schwarzschild radii}, \emph{Astron.
  Astroph.} 595:\penalty0 A54  (Oct., 2016).
\newblock \doi{10.1051/0004-6361/201628829}.

\bibitem{bro+15}
A.~E. {Broderick}, R.~{Narayan}, J.~{Kormendy}, E.~S. {Perlman}, M.~J. {Rieke},
  and S.~S. {Doeleman}, {The Event Horizon of M87}, \emph{Astrophys. J.}
  805:\penalty0 179  (June, 2015).
\newblock \doi{10.1088/0004-637X/805/2/179}.

\bibitem{bla+82}
R.~D. {Blandford} and D.~G. {Payne}, {Hydromagnetic flows from accretion discs
  and the production of radio jets}, \emph{Mon. Not. R. Astron. Soc.} {\bf
  199}, \penalty0 883--903  (June, 1982).
\newblock \doi{10.1093/mnras/199.4.883}.

\bibitem{spr10}
H.~C. {Spruit}.
\newblock {Theory of Magnetically Powered Jets}.
\newblock In ed. T.~{Belloni}, \emph{Lecture Notes in Physics, Berlin Springer
  Verlag}, vol. 794, \emph{Lecture Notes in Physics, Berlin Springer Verlag},
  p. 233  (Mar., 2010).
\newblock \doi{10.1007/978-3-540-76937-8_9}.

\bibitem{cap+08}
S.~{Capozziello}, C.~{Corda}, and M.~F. {de Laurentis}, {Massive gravitational
  waves from $f(R)$ theories of gravity: Potential detection with LISA},
  \emph{Phys. Lett. B}. {\bf 669}, \penalty0 255--259  (Nov., 2008).
\newblock \doi{10.1016/j.physletb.2008.10.001}.

\bibitem{lia+17}
D.~{Liang}, Y.~{Gong}, S.~{Hou}, and Y.~{Liu}, {Polarizations of gravitational
  waves in $f(R)$ gravity}, \emph{Phys. Rev. D}. 95\penalty0 (10):\penalty0
  104034  (May, 2017).
\newblock \doi{10.1103/PhysRevD.95.104034}.

\bibitem{cor18}
C.~{Corda}, {The future of gravitational theories in the era of the
  gravitational wave astronomy}, \emph{Int. J. Mod. Phys. D}. 27:\penalty0
  18500060  (Jan., 2018).

\bibitem{abb+18}
B.~P. {Abbott}, R.~{Abbott}, T.~D. {Abbott}, F.~{Acernese}, K.~{Ackley},
  C.~{Adams}, T.~{Adams}, P.~{Addesso}, R.~X. {Adhikari}, V.~B. {Adya}, and
  et~al., {First Search for Nontensorial Gravitational Waves from Known
  Pulsars}, \emph{Phys. Rev. Lett.} 120\penalty0 (3):\penalty0 031104  (Jan.,
  2018).
\newblock \doi{10.1103/PhysRevLett.120.031104}.

\bibitem{abb+18a}
{The LIGO Scientific Collaboration}, {the Virgo Collaboration}, B.~P. {Abbott},
  R.~{Abbott}, T.~D. {Abbott}, F.~{Acernese}, K.~{Ackley}, C.~{Adams},
  T.~{Adams}, P.~{Addesso}, and et~al., {A Search for Tensor, Vector, and
  Scalar Polarizations in the Stochastic Gravitational-Wave Background},
  \emph{ArXiv e-prints}  (Feb., 2018).

\bibitem{cap+18}
C.~{Caprini} and D.~G. {Figueroa}, {Cosmological Backgrounds of Gravitational
  Waves}, \emph{ArXiv e-prints}  (Jan., 2018).

\bibitem{vai+17}
J.~{Vainio} and I.~{Vilja}, {$f(R)$ gravity constraints from gravitational
  waves}, \emph{Gen. Relativ. Gravit.} 49:\penalty0 99  (Aug., 2017).
\newblock \doi{10.1007/s10714-017-2262-3}.

\bibitem{hu+07}
W.~{Hu} and I.~{Sawicki}, {Models of $f(R)$ cosmic acceleration that evade
  solar system tests}, \emph{Phys. Rev. D}. 76\penalty0 (6):\penalty0 064004
  (Sept., 2007).
\newblock \doi{10.1103/PhysRevD.76.064004}.

\bibitem{sta07}
A.~A. {Starobinsky}, {Disappearing cosmological constant in $f(R)$ gravity},
  \emph{Soviet Journal of Experimental and Theoretical Physics Letters}. {\bf
  86}, \penalty0 157--163  (Oct., 2007).
\newblock \doi{10.1134/S0021364007150027}.

\bibitem{lee17b}
S.~{Lee}, {Constraint on reconstructed $f(R)$ gravity models from gravitational
  waves}, \emph{ArXiv e-prints}  (Nov., 2017).

\bibitem{lee17a}
S.~{Lee}, {Reconstruction of $f(R)$ gravity models from observations},
  \emph{ArXiv e-prints}  (Oct., 2017).

\bibitem{abb+17c}
B.~P. {Abbott}, R.~{Abbott}, T.~D. {Abbott}, F.~{Acernese}, K.~{Ackley},
  C.~{Adams}, T.~{Adams}, P.~{Addesso}, R.~X. {Adhikari}, V.~B. {Adya}, and
  et~al., {Gravitational Waves and Gamma-Rays from a Binary Neutron Star
  Merger: GW170817 and GRB 170817A}, \emph{Astrophys. J. Lett.} 848:\penalty0
  L13  (Oct., 2017).
\newblock \doi{10.3847/2041-8213/aa920c}.

\bibitem{abb+17d}
B.~P. {Abbott}, R.~{Abbott}, T.~D. {Abbott}, F.~{Acernese}, K.~{Ackley},
  C.~{Adams}, T.~{Adams}, P.~{Addesso}, R.~X. {Adhikari}, V.~B. {Adya}, and
  et~al., {Multi-messenger Observations of a Binary Neutron Star Merger},
  \emph{Astrophys. J. Lett.} 848:\penalty0 L12  (Oct., 2017).
\newblock \doi{10.3847/2041-8213/aa91c9}.

\bibitem{ezq+17}
J.~M. {Ezquiaga} and M.~{Zumalac{\'a}rregui}, {Dark Energy After GW170817: Dead
  Ends and the Road Ahead}, \emph{Phys. Rev. Lett.} 119\penalty0 (25):\penalty0
  251304  (Dec., 2017).
\newblock \doi{10.1103/PhysRevLett.119.251304}.

\bibitem{par+18}
K.~{Pardo}, M.~{Fishbach}, D.~E. {Holz}, and D.~N. {Spergel}, {Limits on the
  number of spacetime dimensions from GW170817}, \emph{ArXiv e-prints}  (Jan.,
  2018).

\bibitem{noj+18}
S.~{Nojiri} and S.~D. {Odintsov}, {Cosmological bound from the neutron star
  merger GW170817 in scalar-tensor and $f(R)$ gravity theories}, \emph{Phys.
  Lett. B}. {\bf 779}, \penalty0 425--429  (Apr., 2018).
\newblock \doi{10.1016/j.physletb.2018.01.078}.

\bibitem{ni17}
A.~{Nishizawa}, {Generalized framework for testing gravity with
  gravitational-wave propagation. I. Formulation}, \emph{ArXiv e-prints}
  (Oct., 2017).

\bibitem{ara+17}
S.~{Arai} and A.~{Nishizawa}, {Generalized framework for testing gravity with
  gravitational-wave propagation. II. Constraints on Horndeski theory},
  \emph{ArXiv e-prints}  (Nov., 2017).

\bibitem{sha64}
I.~I. {Shapiro}, {Fourth Test of General Relativity}, \emph{Phys. Rev. Lett.}
  {\bf 13}, \penalty0 789--791  (Dec., 1964).
\newblock \doi{10.1103/PhysRevLett.13.789}.

\bibitem{kah+07}
E.~O. {Kahya} and R.~P. {Woodard}, {A generic test of modified gravity models
  which emulate dark matter}, \emph{Phys. Lett. B}. {\bf 652}, \penalty0
  213--216  (Sept., 2007).
\newblock \doi{10.1016/j.physletb.2007.07.029}.

\bibitem{bor+18}
S.~{Boran}, S.~{Desai}, E.~O. {Kahya}, and R.~P. {Woodard}, {GW170817 falsifies
  dark matter emulators}, \emph{Phys. Rev. D}. 97\penalty0 (4):\penalty0 041501
   (Feb., 2018).
\newblock \doi{10.1103/PhysRevD.97.041501}.

\bibitem{gre+18}
M.~A. {Green}, J.~W. {Moffat}, and V.~T. {Toth}, {Modified gravity (MOG), the
  speed of gravitational radiation and the event GW170817/GRB170817A},
  \emph{Phys. Lett. B}. {\bf 780}, \penalty0 300--302  (May, 2018).
\newblock \doi{10.1016/j.physletb.2018.03.015}.

\bibitem{mof16}
J.~W. {Moffat}, {LIGO GW150914 and GW151226 gravitational wave detection and
  generalized gravitation theory (MOG)}, \emph{Phys. Lett. B}. {\bf 763},
  \penalty0 427--433  (Dec., 2016).
\newblock \doi{10.1016/j.physletb.2016.10.082}.

\bibitem{mof17}
J.~W. {Moffat}, {Misaligned Spin Merging Black Holes in Modified Gravity
  (MOG)}, \emph{ArXiv e-prints}  (June, 2017).

\bibitem{abb+16b}
B.~P. {Abbott}, R.~{Abbott}, T.~D. {Abbott}, M.~R. {Abernathy}, F.~{Acernese},
  K.~{Ackley}, C.~{Adams}, T.~{Adams}, P.~{Addesso}, R.~X. {Adhikari}, and
  et~al., {Tests of General Relativity with GW150914}, \emph{Phys. Rev. Lett.}
  116\penalty0 (22):\penalty0 221101  (June, 2016).
\newblock \doi{10.1103/PhysRevLett.116.221101}.

\bibitem{man+18}
L.~{Manfredi}, J.~{Mureika}, and J.~{Moffat}, {Quasinormal modes of modified
  gravity (MOG) black holes}, \emph{Phys. Lett. B}. {\bf 779}, \penalty0
  492--497  (Apr., 2018).
\newblock \doi{10.1016/j.physletb.2017.11.006}.

\bibitem{wei+18}
S.-W. {Wei} and Y.-X. {Liu}, {Merger estimates for rotating Kerr-MOG black
  holes in modified gravity}, \emph{ArXiv e-prints}  (Mar., 2018).

\bibitem{bha+17}
S.~{Bhattacharyya} and S.~{Shankaranarayanan}, {Quasinormal modes as a
  distinguisher between general relativity and $f(R)$ gravity}, \emph{Phys.
  Rev. D}. 96\penalty0 (6):\penalty0 064044  (Sept., 2017).
\newblock \doi{10.1103/PhysRevD.96.064044}.

\bibitem{naf+11}
J.~{N{\"a}f} and P.~{Jetzer}, {Gravitational radiation in quadratic $f(R)$
  gravity}, \emph{Phys. Rev. D}. 84\penalty0 (2):\penalty0 024027  (July,
  2011).
\newblock \doi{10.1103/PhysRevD.84.024027}.

\bibitem{rom+2013}
G.~E. {Romero}, {Adversus singularitates: The ontology of space-time
  singularities}, \emph{Found. Sci}. \penalty0 (2), \penalty0 297--306  (Oct.,
  2013).

\bibitem{Olm+15b}
G.~{Olmo} and D.~{Rubiera-Garcia}, {Nonsingular Black Holes in $f(R)$
  Theories}, \emph{Universe}. {\bf 1}, \penalty0 173--185  (Aug., 2015).
\newblock \doi{10.3390/universe1020173}.

\bibitem{Olm+16}
G.~J. {Olmo}, D.~{Rubiera-Garcia}, and A.~{Sanchez-Puente}, {Classical
  resolution of black hole singularities via wormholes}, \emph{Eur. Phys. J.
  C}. 76:\penalty0 143  (Mar., 2016).
\newblock \doi{10.1140/epjc/s10052-016-3999-7}.

\bibitem{Bej+17}
C.~{Bejarano}, G.~J. {Olmo}, and D.~{Rubiera-Garcia}, {What is a singular black
  hole beyond general relativity?}, \emph{Phys. Rev. D}. 95\penalty0
  (6):\penalty0 064043  (Mar., 2017).
\newblock \doi{10.1103/PhysRevD.95.064043}.

\bibitem{Per+14}
D.~{P{\'e}rez}, G.~E. {Romero}, and S.~E. {Perez-Bergliaffa}, {An Analysis of a
  Regular Black Hole Interior Model}, \emph{Int. J. Theor. Phys.} {\bf 53},
  \penalty0 734  (Mar., 2014).
\newblock \doi{10.1007/s10773-013-1861-3}.

\bibitem{mof+18}
J.~W. {Moffat}, {Regular Rotating MOG Dark Compact Object}, \emph{ArXiv
  e-prints}  (June, 2018).

\bibitem{Saf02}
M.~{Safonova}, D.~F. {Torres}, and G.~E. {Romero}, {Microlensing by natural
  wormholes: Theory and simulations}, \emph{Phys. Rev. D}. 65\penalty0
  (2):\penalty0 023001  (Jan., 2002).
\newblock \doi{10.1103/PhysRevD.65.023001}.

\bibitem{Barr+09}
C.~{Barrag{\'a}n}, G.~J. {Olmo}, and H.~{Sanchis-Alepuz}, {Bouncing cosmologies
  in Palatini $f(R)$ gravity}, \emph{Phys. Rev. D}. 80\penalty0 (2):\penalty0
  024016  (July, 2009).
\newblock \doi{10.1103/PhysRevD.80.024016}.

\bibitem{Paul+14}
N.~{Paul}, S.~{Nil Chakrabarty}, and K.~{Bhattacharya}, {Cosmological bounces
  in spatially flat FRW spacetimes in metric $f(R)$ gravity}, \emph{J. Cosm.
  Astropart. Phys.} 10:\penalty0 009  (Oct., 2014).
\newblock \doi{10.1088/1475-7516/2014/10/009}.

\bibitem{Bat+16}
K.~{Bhattacharya} and S.~{Chakrabarty}, {Intricacies of cosmological bounce in
  polynomial metric $f(R)$ gravity for flat FLRW spacetime}, \emph{J. Cosm.
  Astropart. Phys.} 2:\penalty0 030  (Feb., 2016).
\newblock \doi{10.1088/1475-7516/2016/02/030}.

\bibitem{QUBIC+16}
J.~{Aumont}, S.~{Banfi}, P.~{Battaglia}, E.~S. {Battistelli}, A.~{Ba{\`u}},
  B.~{B{\'e}lier}, D.~{Bennett}, L.~{Berg{\'e}}, J.~P. {Bernard},
  M.~{Bersanelli}, M.~A. {Bigot-Sazy}, N.~{Bleurvacq}, G.~{Bordier},
  J.~{Brossard}, E.~F. {Bunn}, D.~{Buzi}, A.~{Buzzelli}, D.~{Cammilleri},
  F.~{Cavaliere}, P.~{Chanial}, C.~{Chapron}, G.~{Coppi}, A.~{Coppolecchia},
  F.~{Couchot}, R.~{D'Agostino}, G.~{D'Alessandro}, P.~{de Bernardis}, G.~{De
  Gasperis}, M.~{De Petris}, T.~{Decourcelle}, F.~{Del Torto}, L.~{Dumoulin},
  A.~{Etchegoyen}, C.~{Franceschet}, B.~{Garcia}, A.~{Gault}, D.~{Gayer},
  M.~{Gervasi}, A.~{Ghribi}, M.~{Giard}, Y.~{Giraud-H{\'e}raud}, M.~{Gradziel},
  L.~{Grandsire}, J.~C. {Hamilton}, D.~{Harari}, V.~{Haynes},
  S.~{Henrot-Versill{\'e}}, N.~{Holtzer}, J.~{Kaplan}, A.~{Korotkov},
  L.~{Lamagna}, J.~{Lande}, S.~{Loucatos}, A.~{Lowitz}, V.~{Lukovic},
  B.~{Maffei}, S.~{Marnieros}, J.~{Martino}, S.~{Masi}, A.~{May},
  M.~{McCulloch}, M.~C. {Medina}, S.~{Melhuish}, A.~{Mennella}, L.~{Montier},
  A.~{Murphy}, D.~{N{\'e}el}, M.~W. {Ng}, C.~{O'Sullivan}, A.~{Paiella},
  F.~{Pajot}, A.~{Passerini}, A.~{Pelosi}, C.~{Perbost}, O.~{Perdereau},
  F.~{Piacentini}, M.~{Piat}, L.~{Piccirillo}, G.~{Pisano}, D.~{Pr{\^e}le},
  R.~{Puddu}, D.~{Rambaud}, O.~{Rigaut}, G.~E. {Romero}, M.~{Salatino},
  A.~{Schillaci}, S.~{Scully}, M.~{Stolpovskiy}, F.~{Suarez}, A.~{Tartari},
  P.~{Timbie}, M.~{Tristram}, G.~{Tucker}, D.~{Vigan{\`o}}, N.~{Vittori},
  F.~{Voisin}, B.~{Watson}, M.~{Zannoni}, and A.~{Zullo}, {QUBIC Technical
  Design Report}, \emph{ArXiv e-prints}  (Sept., 2016).

\bibitem{Bou+13}
M.~{Bouhmadi-Lopez}, J.~{Morais}, and A.~B. {Henriques}, {The spectrum of
  gravitational waves in an $f(R)$ model with a bounce}, \emph{ArXiv e-prints}
  (Feb., 2013).

\bibitem{Jam+18}
S.~{Jamali}, M.~{Roshan}, and L.~{Amendola}, {On the cosmology of
  scalar-tensor-vector gravity theory}, \emph{J. Cosm. Astropart. Phys.}
  1:\penalty0 048  (Jan., 2018).
\newblock \doi{10.1088/1475-7516/2018/01/048}.

\bibitem{Arz+18}
Z.~{Arzoumanian}, P.~T. {Baker}, A.~{Brazier}, S.~{Burke-Spolaor}, S.~J.
  {Chamberlin}, S.~{Chatterjee}, B.~{Christy}, J.~M. {Cordes}, N.~J. {Cornish},
  F.~{Crawford}, H.~{Thankful Cromartie}, K.~{Crowter}, M.~{DeCesar}, P.~B.
  {Demorest}, T.~{Dolch}, J.~A. {Ellis}, R.~D. {Ferdman}, E.~{Ferrara}, W.~M.
  {Folkner}, E.~{Fonseca}, N.~{Garver-Daniels}, P.~A. {Gentile}, R.~{Haas},
  J.~S. {Hazboun}, E.~A. {Huerta}, K.~{Islo}, G.~{Jones}, M.~L. {Jones}, D.~L.
  {Kaplan}, V.~M. {Kaspi}, M.~T. {Lam}, T.~J.~W. {Lazio}, L.~{Levin}, A.~N.
  {Lommen}, D.~R. {Lorimer}, J.~{Luo}, R.~S. {Lynch}, D.~R. {Madison}, M.~A.
  {McLaughlin}, S.~T. {McWilliams}, C.~M.~F. {Mingarelli}, C.~{Ng}, D.~J.
  {Nice}, R.~S. {Park}, T.~T. {Pennucci}, N.~S. {Pol}, S.~M. {Ransom}, P.~S.
  {Ray}, A.~{Rasskazov}, X.~{Siemens}, J.~{Simon}, R.~{Spiewak}, I.~H.
  {Stairs}, D.~R. {Stinebring}, K.~{Stovall}, J.~{Swiggum}, S.~R. {Taylor},
  M.~{Vallisneri}, R.~{van Haasteren}, S.~{Vigeland}, W.~W. {Zhu}, and {The
  NANOGrav Collaboration}, {The NANOGrav 11 Year Data Set: Pulsar-timing
  Constraints on the Stochastic Gravitational-wave Background},
  \emph{Astrophys. J.} 859:\penalty0 47  (May, 2018).
\newblock \doi{10.3847/1538-4357/aabd3b}.

\bibitem{Ell+14}
J.~{Ellis}.
\newblock {NANOGrav Limits on Continuous Gravitational Waves from Supermassive
  Black Hole Binaries}.
\newblock In \emph{APS Meeting Abstracts}, p. C15.008  (Mar., 2014).
\newblock \doi{10.1103/BAPS.2014.APRIL.C15.8}.

\end{thebibliography}

\end{document}